\documentclass[twocolumn,prb,amsmath,showpacs,amssymb,amsfonts,10pt]{revtex4-1}

\usepackage{graphicx}
\usepackage{subfigure}
\usepackage[english]{babel}
\usepackage{hyperref}

\begin{document}
\def\bfs{\mathbf{S}}
\def\bfk{\mathbf{k}}
\def\bft{\mathbf{t}}
\def\bfe{\mathbf{e}}
\def\bfx{\mathbf{x}}
\def\bfy{\mathbf{y}}
\def\tr{\mathrm{tr}}

\title{
Interaction of domain walls and vortices in the two-dimensional O(2) and O(3) principal chiral models
}
\author{A.O.~Sorokin}
\email{aosorokin@gmail.com}
\affiliation{Petersburg Nuclear Physics Institute, NRC Kurchatov Institute,\\
188300 Orlova Roscha, Gatchina, Russia}

\date{\today}

\begin{abstract}
Using extensive Monte Carlo simulations, we investigate the critical properties of domain walls, vortices and $\mathbb{Z}_2$ vortices in the Ising-$O(2)$ and Ising-$O(3)\otimes O(2)$ models. We have consider the nontrivial case when disorder in the Ising order parameter induces disorder in the continuous parameter. Such a situation arises when a domain wall becomes opaque for continuous parameter correlations. We find that in this case the vortex density at the BKT transition (or crossover) point turns out to be non-universal, while the wall density at the Ising transition remains universal, i.e. in agreement with the Ising model. An important part of this study is the numerical measurement of defect-defect correlators. We find that the wall-vortex correlator tends to zero in the thermodynamic limit at the Ising point, which explains the universality of the wall density. A possible multicritical behavior of the models is also discussed.
\end{abstract}


\maketitle

\section{Introduction}

In our previous paper \cite{Sorokin17}, we have considered a lattice version of the $O(3)$ principal chiral model (or equivalently the $O(3)\otimes O(3)$ sigma model) in two dimensions. This model is interesting at least in that it describes frustrated magnets with non-planar spin orderings. Another reason to explore this model is that it contains two types of topological defects: domain walls corresponding to the discrete part of the order parameter space $G/H=O(3)\equiv\mathbb{Z}_2\otimes SO(3)$, and $\mathbb{Z}_2$ vortices arising due to the continuous part is not simply-connected, with the fundamental group $\pi_1(SO(3))=\mathbb{Z}_2$. In contrast to usual vortices of the $O(2)$ model, $\mathbb{Z}_2$ vortices associating in pairs at low temperature do not lead to appearance of a quasi-long-range order and to occurrence of a Berezinskii-Kosterlitz-Thouless (BKT) phase transition. But the increase in the vortex density leads to a rather sharp change of the temperature behavior from one predicted by the $O(4)$ sigma model at low temperature to some high-temperature one in the form of an explicit crossover\cite{Southern95,Azaria01}. Moreover, $\mathbb{Z}_2$ vortices are responsible for a change of type of the phase transition in the discrete order parameter of the $O(3)$ principal chiral model: instead of a Ising-like second-order transition, one observes a first-order one. The role of vortices in critical behavior is revealed by the fact that the transition and crossover occur at the same temperature, and the densities of domain walls and vortices have a jump at the critical point\cite{Sorokin17}.

As it has been noted in ref.\cite{Domenge08}, if $\mathbb{Z}_2$ vortices are very heavy, and the vortex density is insignificant at the temperature of the transition in the discrete order parameter, one expects that this transition is of second-order and falls in the 2D Ising universality class. It means that the crossover temperature is larger than the transition temperature $T_{\mathbb{Z}_2}>T_\mathrm{dw}$. One can propose the opposite situation when domain walls are heavy, and the transition and crossover occur separately in temperature too but with $T_{\mathbb{Z}_2}<T_\mathrm{dw}$. The critical behavior at the transition in the discrete order parameter in this case remains unexplored until recently\cite{Sorokin18}.

Even if the transition and crossover occur at different temperatures, it does not mean that an interaction between domain walls and vortices becomes inessential. So, one should expect that this interaction may affect the critical behavior. Moreover, as we show in this paper, this interaction may determine a sequence of transitions.

Similar situation with two types of topological defects, namely domain walls and vortices, has been investigated in the context of the Ising-XY model \cite{Granato87,Granato91,Granato91-2,Granato95,Hasenbusch05,Hasenbusch05-2} with the Hamiltonian
\begin{equation}
  H=-\sum_{ij} \left((J+J_B\sigma_i\sigma_j)\bfs_{i}\bfs_{j}+J_C\sigma_i\sigma_j\right),
\end{equation}
\begin{figure}[t]
\center
\includegraphics[scale=0.45]{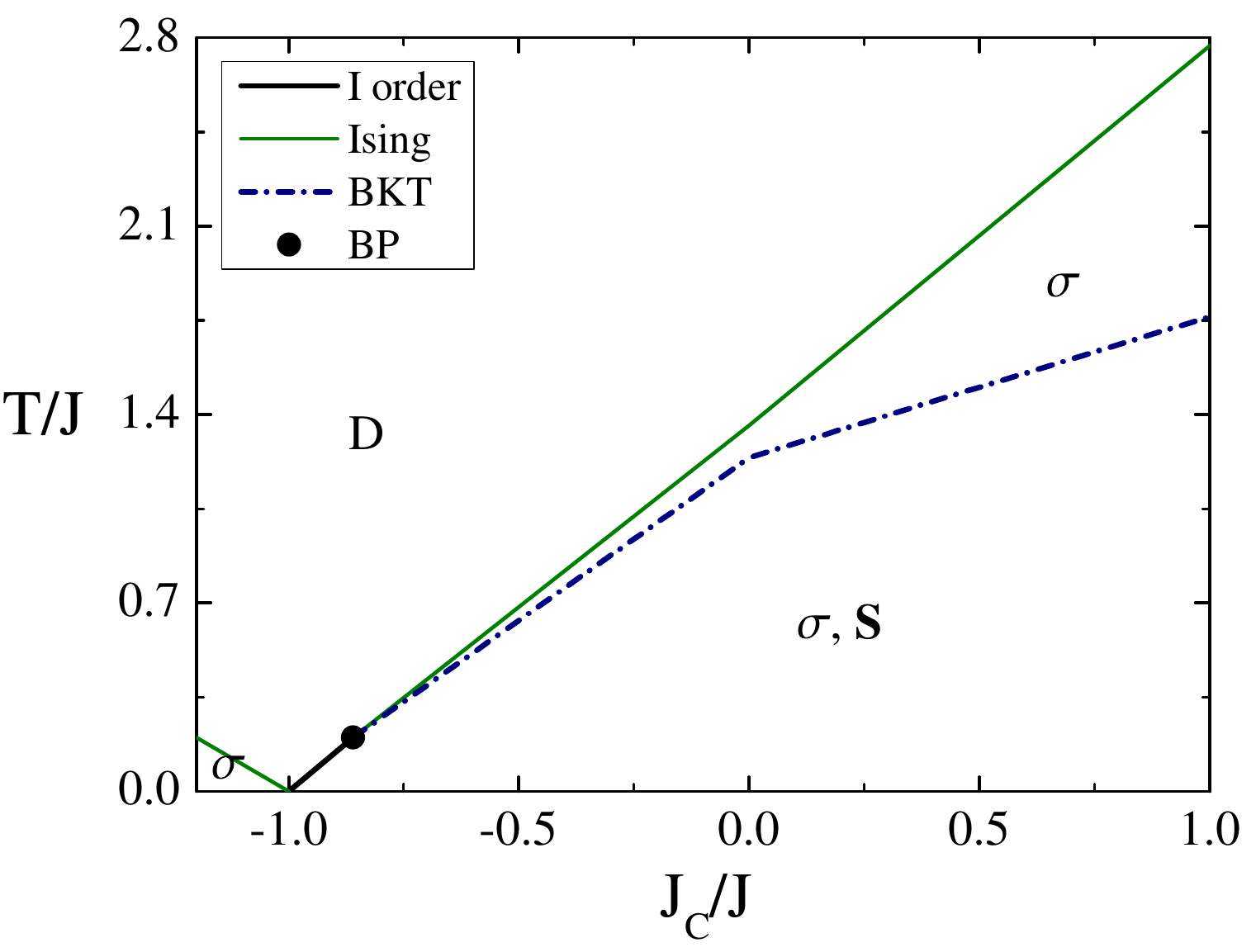}
\caption{\label{fig1} Sketch of the phase diagram of the Ising-XY model at $J_B=J$. Letters $\sigma$ and $\bfs$ mark corresponding long-range and quasi-long-range orders, $D$ marks the disordered phase, BP is the bifurcation point.}
\end{figure}%
where $\bfs_{i}=(\cos\varphi_i,\sin\varphi_i)$ is a two-component ($N=2$) classical vector, $\sigma_i=\pm1$, and the sum $ij$ runs over neighboring sites of a square lattice. The case $J_B=J$ is the most interesting because here the Ising disorder induces the XY disorder due to a domain wall makes XY spins decoupled at a wall: $J+J_B\sigma_i\sigma_j=0$. Thus, one finds that the BKT transition temperature cannot be larger than the Ising transition temperature, $T_\mathrm{v}\leq T_\mathrm{dw}$. Above the value $J_C^B$ corresponding to the bifurcation point, the transition temperatures are different and $T_\mathrm{v}< T_\mathrm{dw}$ (see fig. \ref{fig1}). Below the the bifurcation point $J_C<J_C^B$, but $J_C>-1$, the Ising and BKT transitions occur at the same temperature as a first-order transition.

The case $J_B=J$ relates to the critical phenomena in the wide class of systems, the ground state of which breaks the symmetry group $G/H=O(2)\equiv\mathbb{Z}_2\otimes SO(2)$ (see ref. \cite{Korshunov06} for a review). Such systems as a Josephson junctions array in a perpendicular magnetic field, triangular XY antiferromagnet\cite{Granato87,Granato91,Granato91-2}, and XY frustrated helimagnet\cite{Sorokin12,Sorokin12-2} fall in this class. An important feature of these and similar systems is the presence in the topological excitation spectrum of fractional vortices, which are corners or kinks of domain walls (depending on a specific model)\cite{Halsey85,Korshunov86,Korshunov86-2,Uimin91,Uimin94}. The effective logarithmical interaction of kinks is weaker than the interaction of conventional vortices and leads to a BKT-like phase transition on a domain wall at $T_\mathrm{fv}<T_\mathrm{v}$. So, at $T>T_\mathrm{fv}$, a domain wall turns opaque for the continuous order parameter correlations, analogously to the case $J_B=J$ of the Ising-XY model. As a consequence, on approaching a Ising-like transition, the quasi-long-range order has to break down, and a BKT transition has to occurs at $T_\mathrm{v}<T_\mathrm{dw}$ \cite{Korshunov02}. The exception is the case $T_\mathrm{fv}=T_\mathrm{dw}$ where both transitions occur at the same temperature as a first-order transition.

Of course, the both phenomena, namely the domain wall opacity for the continuous parameter correlations and presence of fractional vortices, are crucial for a sequence of phase transitions as well as for the critical behavior upon them. As far as fractional vortices are absent in the Ising-XY model and have no analogue for $\mathbb{Z}_2$ vortices, in the present work we consider only the first phenomenon and its influence to the critical behavior in the case of two transitions separated in temperature.

To investigate an interaction of domain walls and $\mathbb{Z}_2$ vortices, we introduce the model similar to the Ising-XY model. The simplest model describing $\mathbb{Z}_2$ vortices is so-called $V_{3,2}$ Stiefel model, which has been introduced in ref.\cite{Zumbach93} and studied in refs.\cite{Sorokin17,Sorokin18}. The model is the lattice realization of the $O(3)\otimes O(2)$ sigma model. Domain walls and $\mathbb{Z}_2$ vortices are presented in the Ising-$V_{3,2}$ model\cite{Sorokin18} with the Hamiltonian
\begin{equation}
  H=-\sum_{ij} \left((J+J_B\sigma_i\sigma_j)\tr\,\Phi_i^T\Phi_j+J_C\sigma_i\sigma_j\right),
\end{equation}
where $\Phi_i=(\bfs_i,\bfk_i)$ is a $3\times 2$ matrix composed of two orthogonal unit $3$-vectors $\bfs_i$ and $\bfk_i$. Again, in the case $J_B=J$, the Ising disorder induces the $SO(3)$ disorder by the same reason as in the Ising-XY model.

In the work \cite{Sorokin18} we have studied the both cases of well-separated transition for the Ising-XY model $T_\mathrm{v}<T_\mathrm{dw}$ and $T_\mathrm{v}>T_\mathrm{dw}$ as well as for the Ising-$V_{3,2}$ model $T_{\mathbb{Z}_2}<T_\mathrm{dw}$ and $T_{\mathbb{Z}_2}>T_\mathrm{dw}$ using correspondingly $J_C\gg0$ and $J_C=0$, $J_B\ll J$. We have found that the critical behavior at transitions of all types (including the crossover) falls distinctively to the corresponding classes of the 2D Ising model, BKT-transition and $O(4)$-$\sigma$-model --- high-temperature crossover.

In the current work, using Monte Carlo simulations, we consider less trivial cases including $J_B=J$ where a wall-vortex interaction is significant, determining a sequence of phase transitions. From simulations, we estimate the wall-vortex correlator, describing the local wall-vortex interaction, and show that this interaction is non-vanishing as long as the densities of domain walls and vortices are non-vanishing too with the one important exception: at the point of the transition in the discrete order parameter the correlator tends to zero value in the thermodynamical limit (the limit of a infinite lattice size $L\to\infty$). Thus, the critical behavior remains universal and falls to the 2D Ising universality class. At the transition (or crossover) induced by vortices, the wall-vortex interaction is non-zero. It does not change the main universal properties of the transition, such as the value of the helicity modulus jump at a BKT transition, but changes the critical properties of vortices, including the critical value of the vortex density.

\section{Models and methods}

\begin{figure}[t]
\center
\includegraphics[scale=0.45]{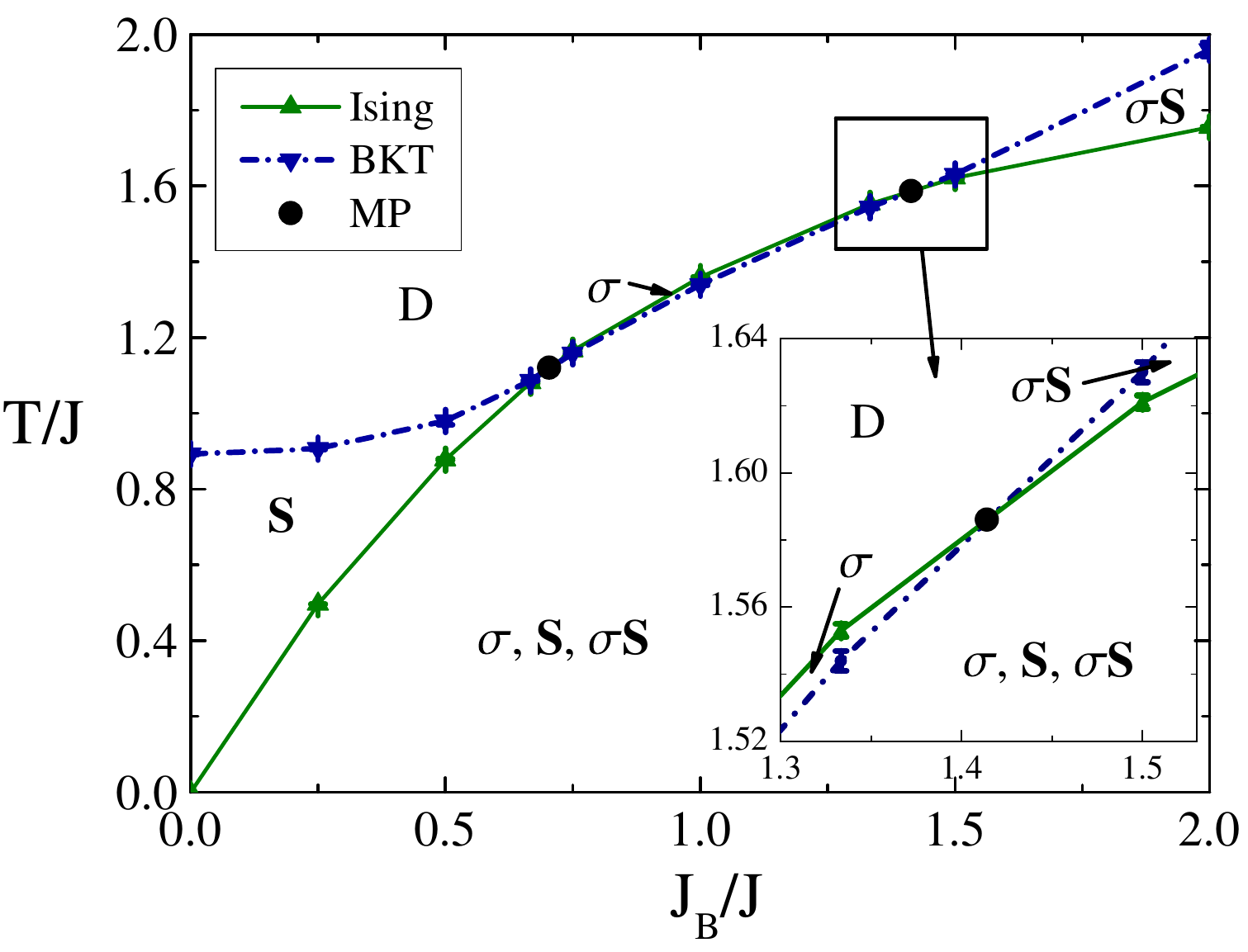}
\caption{\label{fig2} Phase diagram of the Ising-XY model at $J_C=0$. Letters $\sigma$, $\bfs$ and $\sigma\bfs$ mark corresponding long-range and quasi-long-range orders, $D$ marks the disordered phase, MP is a multicritical point.}
\end{figure}%
To study the Ising-XY and Ising-$V_{3,2}$ models, we use Monte Carlo simulations based on the over-relaxed algorithm \cite{Brown87,Creutz87}. We consider a square lattice with periodic boundary conditions. Lattice sizes are $L=24,\,36,\,48,\,60,$ and 90. Thermalization is performed within $3\cdot10^5$ Monte Carlo steps per spin, and calculation of averages within $2.4\cdot10^6$ steps. Details of transition temperature estimation methods for all types of transitions have been discussed in ref.\cite{Sorokin18}. For an Ising-like transition, we use Binder cumulant crossing method \cite{Binder81}. For a BKT-like transition, the Weber-Minnhagen finite-size scaling method \cite{Weber88} has been used. Finally, a crossover temperature is estimated as a value at which the helicity modulus dependence on a lattice size $L$ ceases to correspond to the $O(4)$ sigma model \cite{Azaria92}.

As we have noted in ref.\cite{Sorokin18} for the Ising-XY, since $\sigma^2=1$, a case $J_B>J$ is dual to the case $J_B<J$ that is carried out transformations $J_B\leftrightarrow J$ and $\bfs\leftrightarrow\sigma\bfs$. So, we monitor the three order parameters: the Ising-like $\sigma$, and two XY-like $\bfs$ and $\bfk=\sigma\bfs$. Relevant order parameters for different values of $J_B$ are shown in fig. \ref{fig2}. As a consequence, we introduce three types of topological defects. The first one is domain walls
\begin{equation}
    \tilde\rho_\mathrm{dw}=\frac{1}{4L^2}\sum_{ij}(1-\sigma_i \sigma_j),
    \quad \rho_\mathrm{dw}=\langle\tilde\rho_\mathrm{dw}\rangle,
    \label{ro-dw-ising}
\end{equation}
where $L^2$ is a lattice volume. The second one is ordinary vortices
\begin{equation}
    \tilde\rho_\mathrm{v}(\bfs)=\frac{1}{2\pi L^2}\sum_{\bfx}\sum_{\square_\bfx}\varphi_{ij},
    \quad \rho_\mathrm{v}(\bfs)=\langle\tilde\rho_\mathrm{v}(\bfs)\rangle,
\end{equation}
where $\bfx$ runs over primitive cells of a lattice, $\square_\bfx$ means coming over a cell and summing differences of spin phases $\varphi_{ij}=\varphi_i-\varphi_j\in(-\pi,\pi]$, $\mathbf{S}=(\cos(\varphi_i),\sin(\varphi_i))$. And the third one is alternatively defined vortices
\begin{equation}
    \tilde\rho_\mathrm{v}(\sigma\bfs)=\frac{1}{2\pi L^2}\sum_{x}\sum_{\square_x}\psi_{ij},
    \quad \rho_\mathrm{v}(\sigma\bfs)=\langle\tilde\rho_\mathrm{v}(\sigma\bfs)\rangle,
\end{equation}
$\psi_{ij}=\psi_i-\psi_j\in(-\pi,\pi]$, $\bfk=(\cos(\psi_i),\sin(\psi_i))$.

\begin{figure}[t]
\center
\includegraphics[scale=0.45]{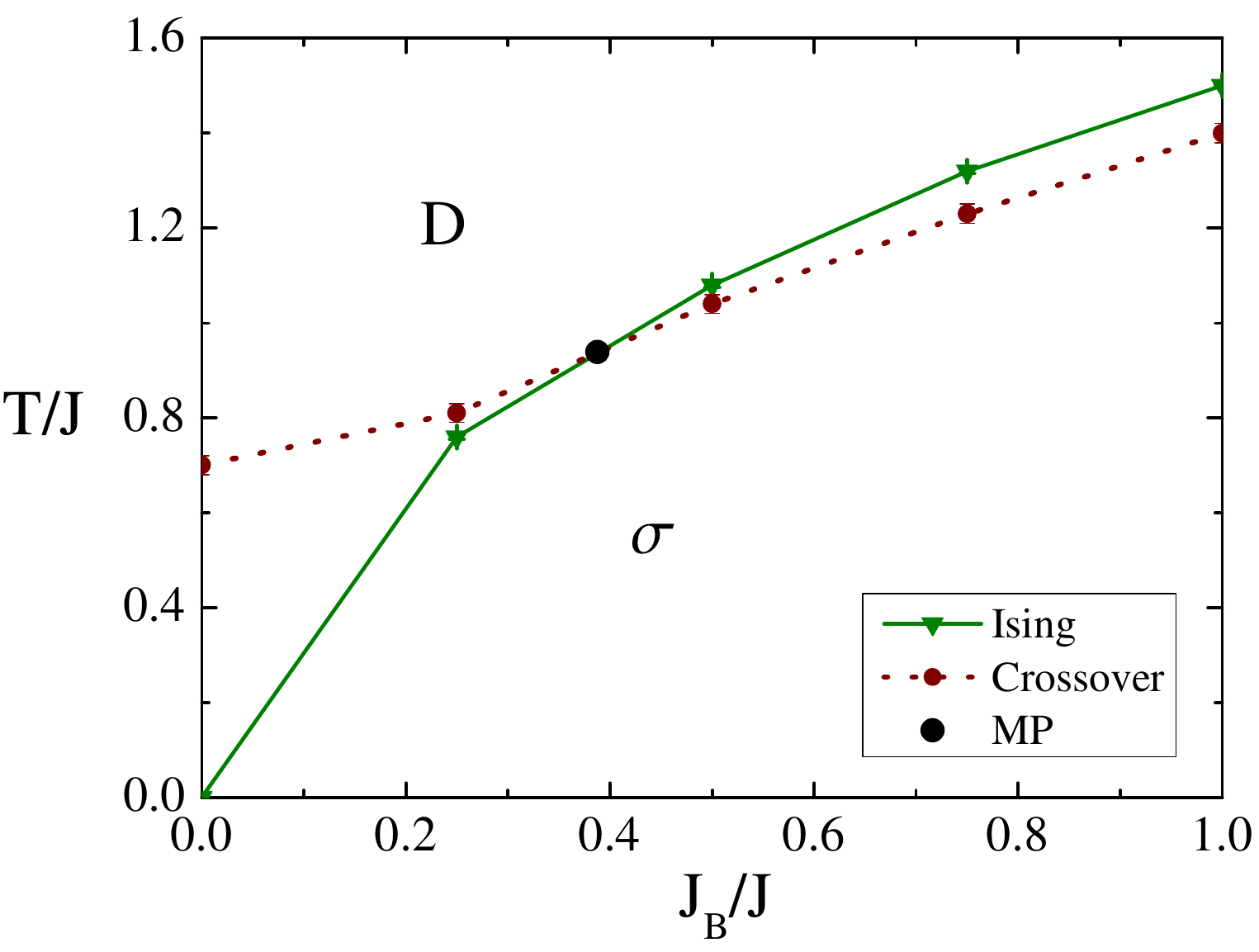}
\caption{\label{fig3} Phase diagram of the Ising-$V_{3,2}$ model at $J_C=0$. MP is the multicritical point.}
\end{figure}%
To define the $\mathbb{Z}_2$ vortex density, we extend the order parameter $\Phi$ of the Ising-$V_{3,2}$ model to a $3\times3$ $SO(3)$ matrix by adding the third vector $\bft=\bfs\times\bfk$, $\det\Phi=1$:
$$
    \tilde\rho_{\mathbb{Z}_2}(\Phi)=\frac{1}{2L^2}\sum_x \left(1-\frac12\tr\prod_{\square_x} f(\Phi_i^{-1}\Phi_j)\right),
$$
\begin{equation}
    \rho_{\mathbb{Z}_2}(\Phi)=\langle\tilde\rho_{\mathbb{Z}_2}(\Phi)\rangle,
\end{equation}
where $f:SO(3)\to SU(2)$ is homomorphism describing double covering of $SO(3)$ by $SU(2)$ constructed using the parametrization of an orthogonal matrix by Euler angles. Also, we introduce the density
$$
    \tilde\rho_{\mathbb{Z}_2}(\sigma\Phi)=\frac{1}{2L^2}\sum_x \left(1-\frac12\tr\prod_{\square_x} f(\Psi_i^{-1}\Psi_j)\right),
$$
\begin{equation}
    \rho_{\mathbb{Z}_2}(\sigma\Phi)=\langle\tilde\rho_{\mathbb{Z}_2}(\sigma\Phi)\rangle,
\end{equation}
where $\Psi_i=(\sigma_i\bfs_i,\sigma_i\bfk_i,\bft_i)$ is also a $SO(3)$ matrix.

\begin{figure}[t]
\center
\includegraphics[scale=0.45]{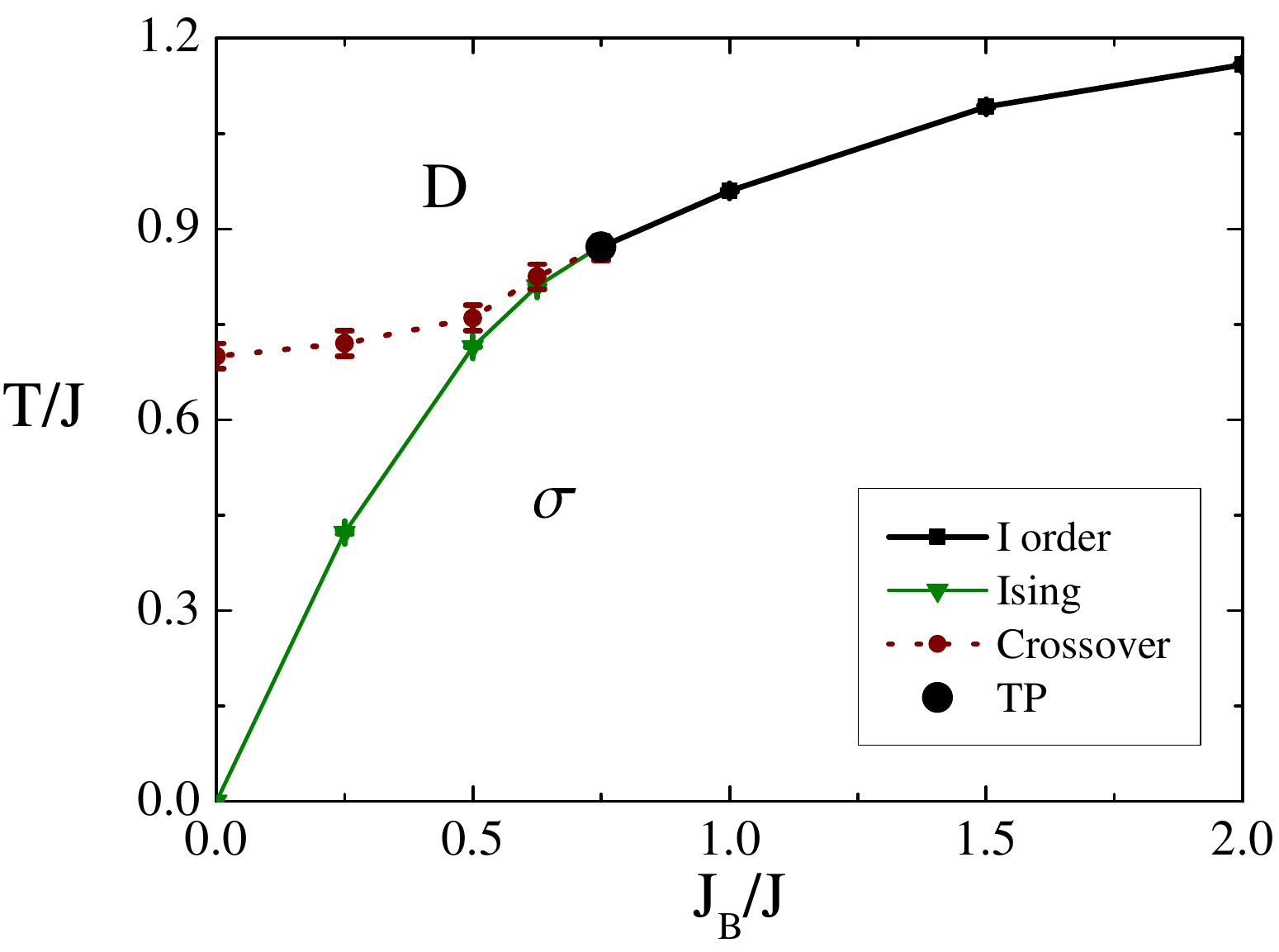}
\caption{\label{fig4} Phase diagram of the interpolating $V_{3,2}$-$V_{3,3}$ model. TP is the tricritical point.}
\end{figure}%
Fig. \ref{fig3} shows the phase diagram of the Ising-$V_{3,2}$ model for $J_C=0$. One can see that the case with well-separated transition temperatures corresponds to too weak coupling value $J_B<0.25J$, therefore we consider one more model called the interpolating $V_{3,2}$-$V_{3,3}$ model \cite{Sorokin18} with the Hamiltonian
\begin{equation}
    H=-\sum_{ij}\left(J\tr\,\Phi_i^T\Phi_j+J_B\sigma_i\sigma_j\bft_i\bft_j\right), \quad \bft_i=\bfs\times\bfk.
    \label{Stiefel}
\end{equation}
So, at the value $J_B=0$, the model is the $V_{3,2}$ Stiefel model, and at $J_B=1$ is the $V_{3,3}$ model \cite{Sorokin17}. Topological defects are determined in the same way as in the Ising-$V_{3,2}$ model. The phase diagram of the interpolating $V_{3,2}$-$V_{3,3}$ model is shown in fig. \ref{fig4}.

The main goal of this paper is to estimate numerically defect-defect correlators. Lets consider two quantities. The first one is the wall-vortex correlator
\begin{equation}
    \langle\rho_\mathrm{dw}\rho_\mathrm{v}\rangle\equiv\frac{1}{4 L^4}
    \left\langle\sum_{\bfx,\bfy}\left|\rho_\mathrm{v}(\bfy)\right|\sum_{\square_\bfx}(1-\sigma_i\sigma_j)\right\rangle,
\end{equation}
where $\rho_\mathrm{v}(\bfy)$ is the local defined topological charge of vortices (both usual or $\mathbb{Z}_2$). The second one is the vortex-vortex correlator
\begin{equation}
    \langle\rho_\mathrm{v_1}\rho_\mathrm{v_2}\rangle\equiv\frac{1}{L^4}
    \left\langle\sum_{\bfx,\bfy}\left|\rho_\mathrm{v_1}(\bfx)\right|\left|\rho_\mathrm{v_2}(\bfy)\right|\right\rangle.
\end{equation}
But we find that these definitions are overabundant due to locality of the defect-defect interactions $\langle\rho_\mathrm{dw}(\bfx)\rho_\mathrm{v}(\bfy)\rangle\sim\delta(\bfx-\bfy)$ and $\langle\rho_\mathrm{v_1}(\bfx)\rho_\mathrm{v_2}(\bfy)\rangle\sim\delta(\bfx-\bfy)$. So the correlators read
\begin{equation}
    \langle\rho_\mathrm{dw}\rho_\mathrm{v}\rangle\equiv\frac{1}{4 L^2}
    \left\langle\sum_{\bfx}\left|\rho_\mathrm{v}(\bfx)\right|\sum_{\square_\bfx}(1-\sigma_i\sigma_j)\right\rangle,
\end{equation}
\begin{equation}
    \langle\rho_\mathrm{v_1}\rho_\mathrm{v_2}\rangle\equiv\frac{1}{L^2}
    \left\langle\sum_{\bfx}\left|\rho_\mathrm{v_1}(\bfx)\right|\left|\rho_\mathrm{v_2}(\bfx)\right|\right\rangle.
\end{equation}
Since $\langle\rho_\mathrm{td}\rangle\neq0$ at non-zero temperature, one should at least consider the covariance $\langle\rho_\mathrm{td_1}\rho_\mathrm{td_2}\rangle-\langle\rho_\mathrm{td_1}\rangle\langle\rho_\mathrm{td_2}\rangle$. But the most informative quantity is the Pearson correlation coefficient
\begin{equation}
    U_{\mathrm{td_1-td_2}}\equiv
    \frac{\langle\rho_\mathrm{td_1}\rho_\mathrm{td_2}\rangle-\langle\rho_\mathrm{td_1}\rangle\langle\rho_\mathrm{td_2}\rangle}{\sqrt{\chi_\mathrm{td_1}\chi_\mathrm{td_2}}},
    \label{td-td-corr}
\end{equation}
where $\chi_\mathrm{td_1}$ is the topological susceptibility
\begin{equation}
\chi_\mathrm{td}=L^2\left(\langle\rho_\mathrm{td}^2\rangle-\langle\rho_\mathrm{td}\rangle^2\right).
\label{chi}
\end{equation}
The cumulant $U$ has the meaning of the effective coupling constant of local defect-defect interactions. As long as the topological defect densities are jointly normally distributed quantities, the vanishing of the cumulant $U$ means that topological defects have no local interactions, both linear or non-linear.

\section{Results}

\begin{figure}[t]
\center
\includegraphics[scale=0.45]{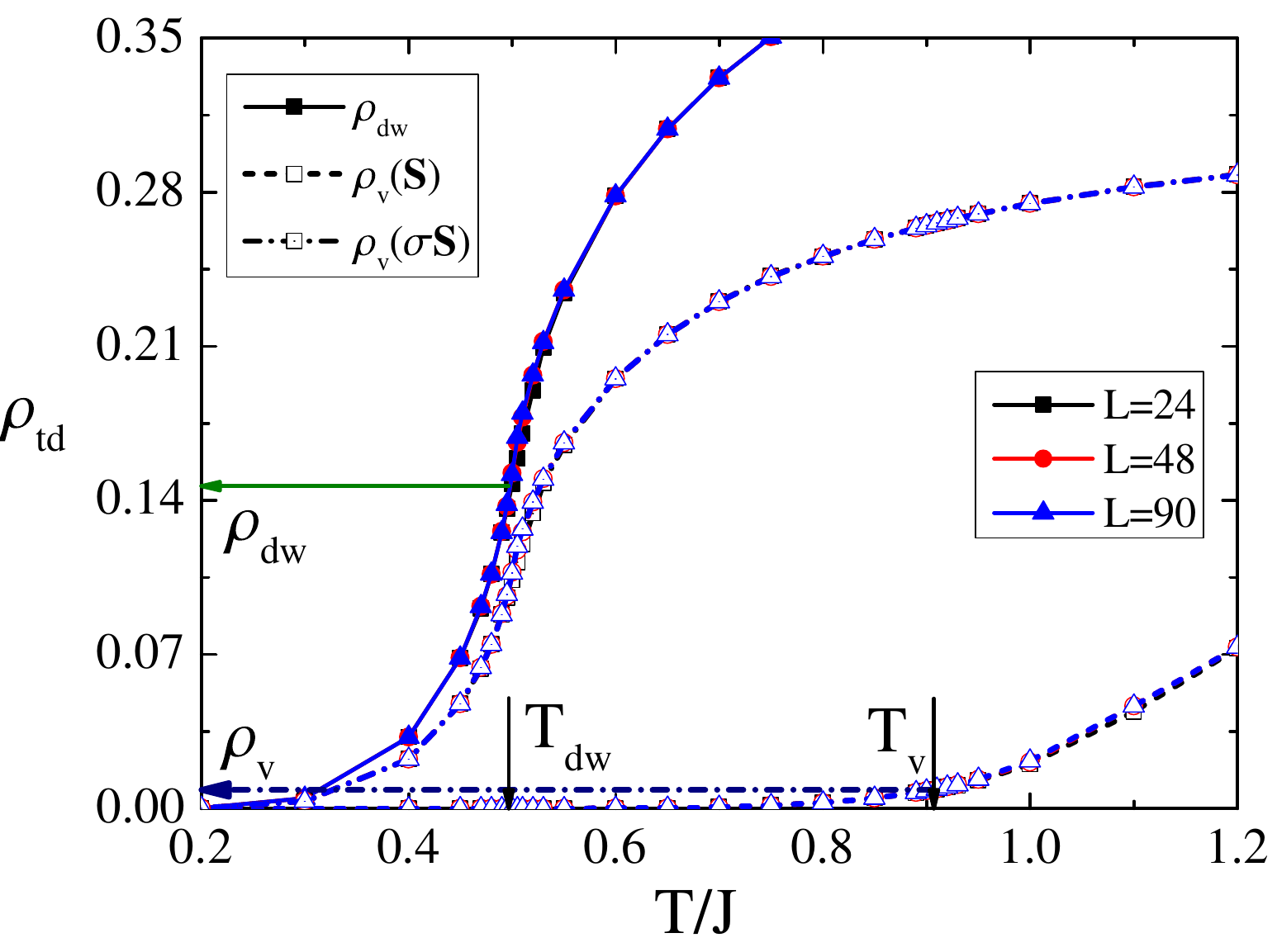}
\caption{\label{fig5} Densities of topological defects in the Ising-XY model at $J_C=0$ and $J_B=0.25$. $\rho_\mathrm{dw}$ and $\rho_\mathrm{v}$ denote the expected values of the critical defect densities at corresponding transitions.}
\end{figure}%
\begin{figure}[t]
\center
\includegraphics[scale=0.45]{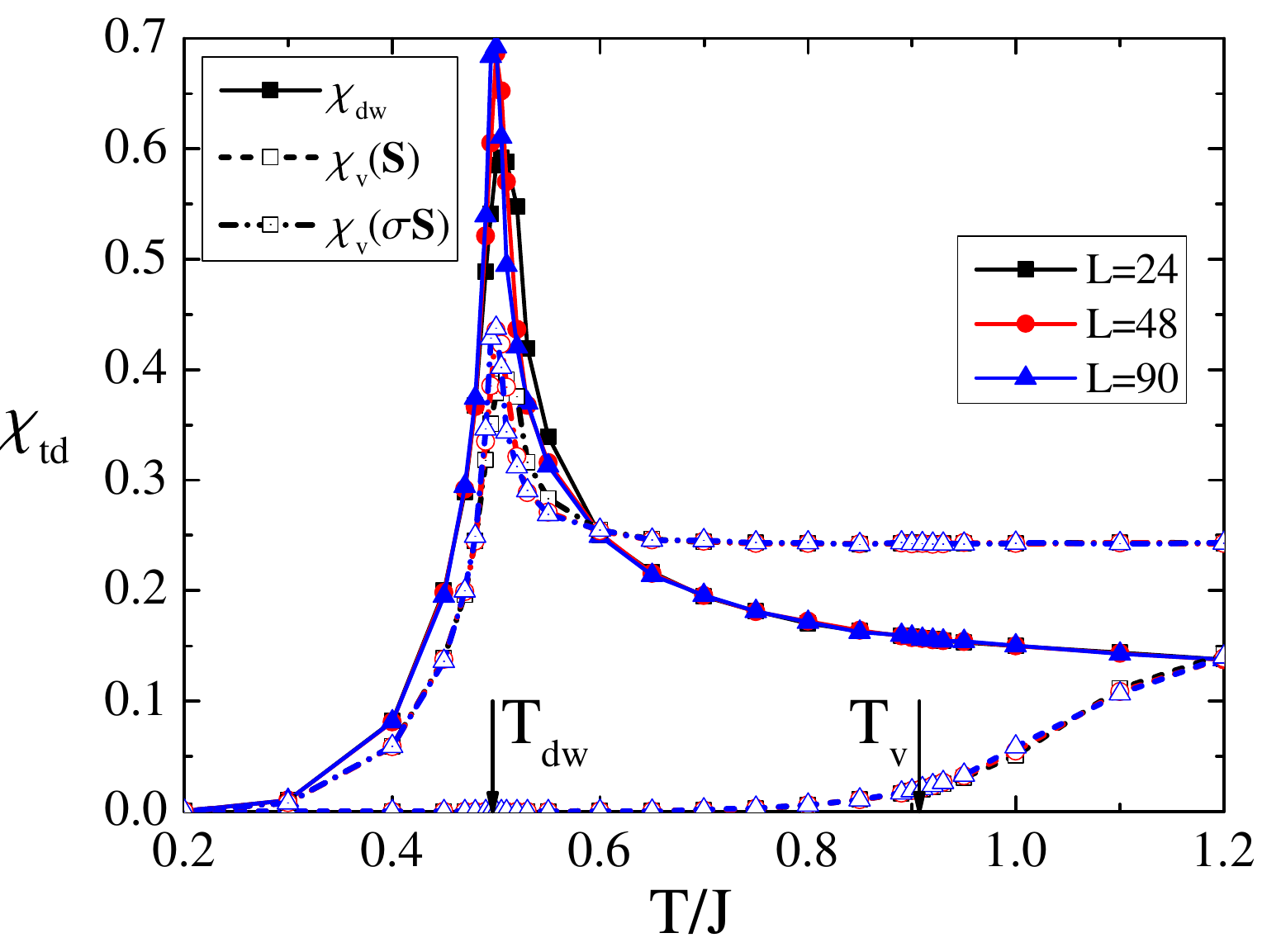}
\caption{\label{fig6} Topological susceptibilities in the Ising-XY model at $J_C=0$ and $J_B=0.25$.}
\end{figure}%
\subsection{Ising-XY model}

In the paper \cite{Sorokin18} we have checked the hypothesis that the density of topological defects has an universal critical value at a transition temperature. In particular, the domain wall density has the value $\rho_\mathrm{dw}\approx0.14644$ at an Ising-like transition point $T=T_\mathrm{dw}$ (for walls determined on a square lattice, see ref. \cite{Sorokin18} for details), and the critical value of the vortex density is $\rho_\mathrm{v}\approx0.0084$ upon a BKT transition $T=T_\mathrm{v}$. Moreover, if the Ising and BKT transitions are well-separated in temperature, we have find the universal values too.

Figs. \ref{fig5} shows the thermal dependence of the topological defect densities of the Ising-XY model at $J_C=0$ and $J_B/J=0.25$. The value of the wall density $\rho_\mathrm{dw}$ at $T=T_\mathrm{dw}$ is close to the universal one as well as the value of the vortex density $\rho_\mathrm{v}(\bfs)$ at $T=T_\mathrm{v}$. The density $\rho_\mathrm{v}(\sigma\bfs)$ is not relevant, it behaves differently than it is expected for the vortex density, while the density $\rho_\mathrm{v}(\bfs)$ behaves similar to the pure $O(2)$ model \cite{Sorokin18} and does not sense the Ising-like transition. The topological susceptibilities $\chi_\mathrm{dw}$ and $\chi_\mathrm{v}(\sigma\bfs)$ has singularities at $T=T_\mathrm{dw}$ (fig. \ref{fig6})
\begin{equation}
    \chi_\mathrm{dw}\sim\chi_\mathrm{v}(\sigma\bfs)\sim A_\mathrm{td}+B_\mathrm{td} \ln L, \quad
    T=T_\mathrm{dw},
\end{equation}
with some constants $A_\mathrm{td}$ and $B_\mathrm{td}$.

\begin{figure}[t]
\center
\includegraphics[scale=0.45]{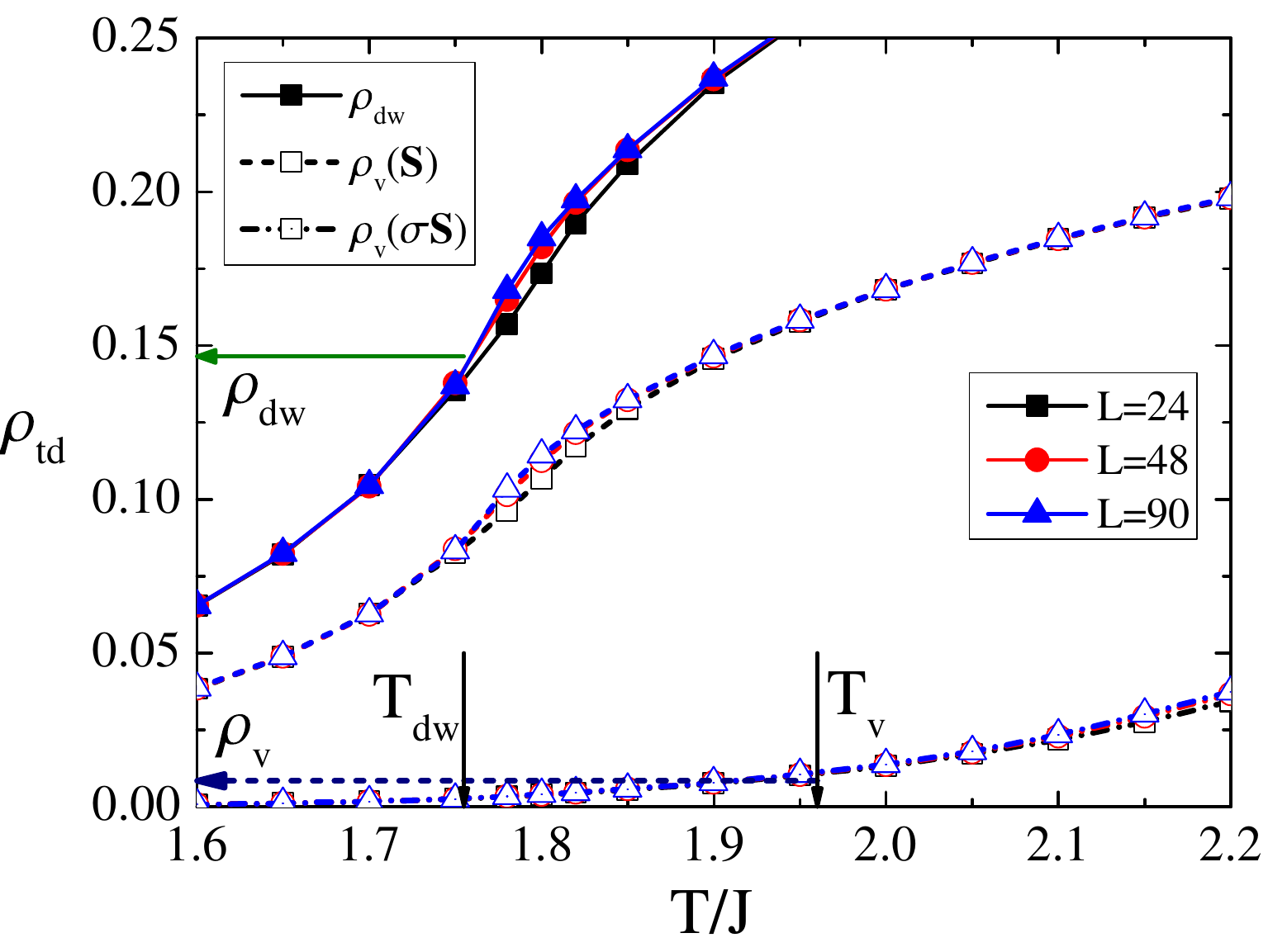}
\caption{\label{fig7} Densities of topological defects in the Ising-XY model at $J_C=0$ and $J_B=2$. $\rho_\mathrm{dw}$ and $\rho_\mathrm{v}$ denote the expected values of the critical defect densities at corresponding transitions.}
\end{figure}%
At $J_C=0$ and $J_B/J=2$, one observe similar picture (fig. \ref{fig7}) with the only difference is that roles of the densities $\rho_\mathrm{v}(\bfs)$ and $\rho_\mathrm{v}(\sigma\bfs)$ are swapped in accordance with the duality. Here, the density $\rho_\mathrm{v}(\sigma\bfs)$ has the typical BKT behavior.

\begin{figure}[t]
\center
\includegraphics[scale=0.45]{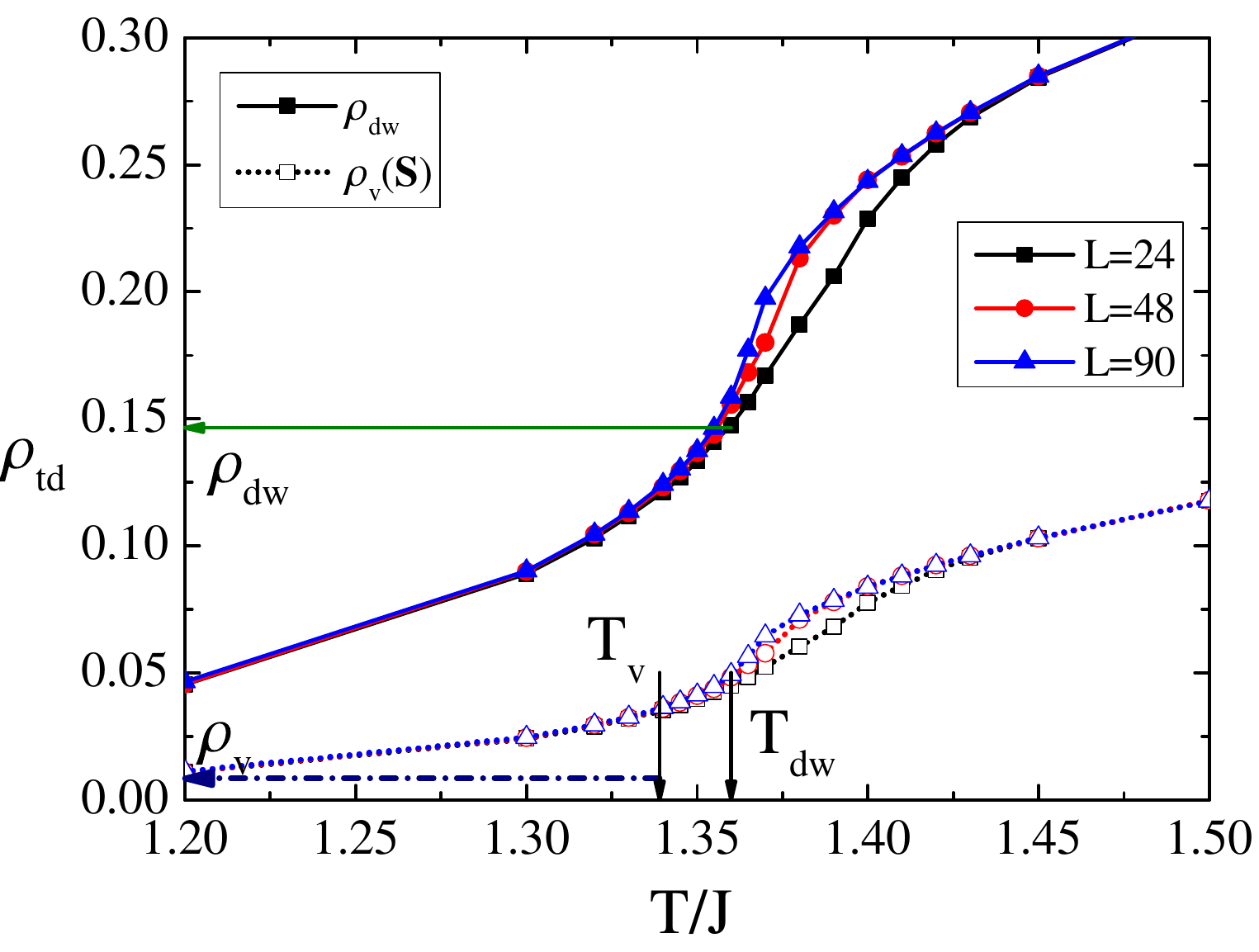}
\caption{\label{fig8} Densities of topological defects in the Ising-XY model at $J_C=0$ and $J_B=1$. $\rho_\mathrm{dw}$ and $\rho_\mathrm{v}$ denote the expected values of the critical defect densities at corresponding transitions.}
\end{figure}%
\begin{figure}[t]
\center
\includegraphics[scale=0.45]{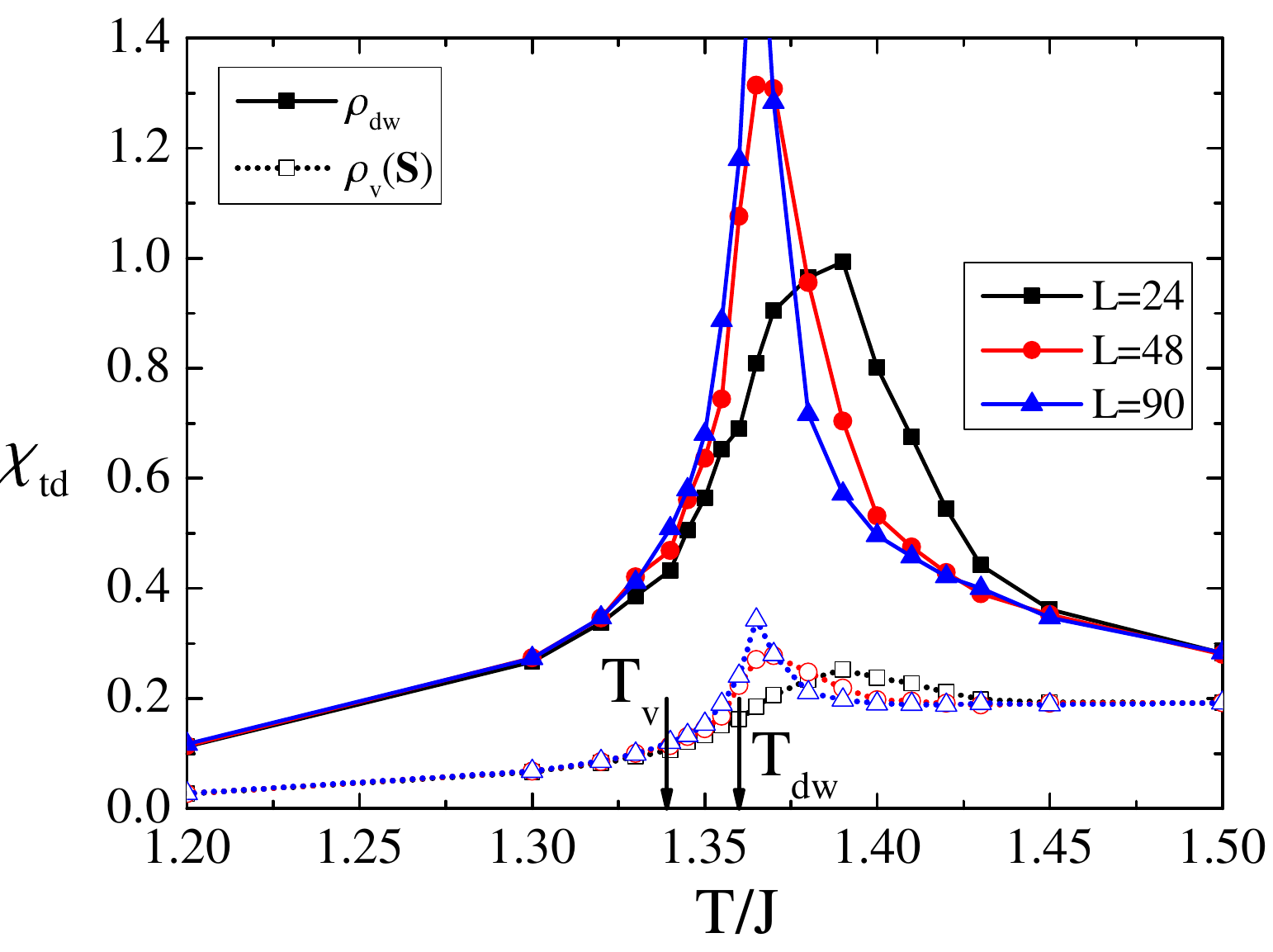}
\caption{\label{fig9} Topological susceptibilities in the Ising-XY model at $J_C=0$ and $J_B=1$.}
\end{figure}%
In ref.\cite{Sorokin18}, the case $J_C=1$ with $T_\mathrm{dw}>T_\mathrm{v}$ has been also considered. It together with two cases considered above relates to the situation with two well-separated in temperature transitions and with the universal values of $\rho_\mathrm{dw}$ at $T_\mathrm{dw}$ and $\rho_\mathrm{v}$ at $T_\mathrm{v}$. At first sight, such a universality is a result of a weakness of correlations between domain walls and the continuous order parameter (including vortices). One can expect that the picture changes in the case $J_B\approx J$, when domain walls induce the XY disorder and an interaction between walls and XY spins (and vortices) can not be ignored. However, fig. \ref{fig8} shows that at least the critical value of the domain wall density $\rho_\mathrm{dw}$ remains universal even at $J_B=J$. Indeed, it is not obvious that an interaction of domain walls with vortices and spin waves determines the sequence of phase transitions but in the same time does not change the critical properties of domain walls.
Howbeit, the critical properties of the vortex density change dramatically. At first, the critical value of the density at the BKT transition $T=T_\mathrm{v}$ is non-universal, and secondly, the vortex density senses the Ising transition and has a singularity at $T=T_\mathrm{dw}$ (fig. \ref{fig9}).

\begin{figure}[t]
\center
\includegraphics[scale=0.45]{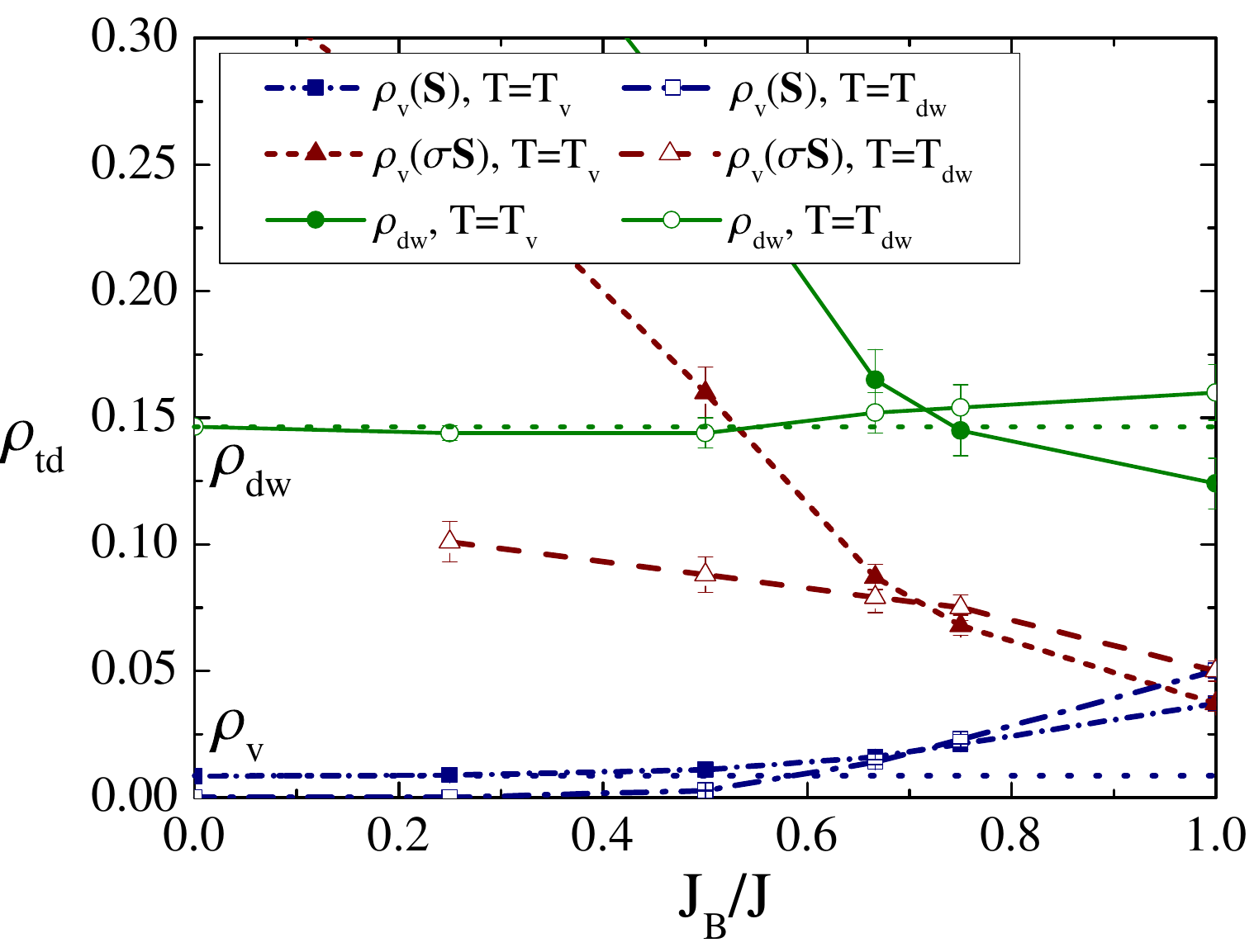}
\caption{\label{fig10} Dependence of the critical value of the defect densities on the parameter $J_B$ in the Ising-XY model at $J_C=0$. $\rho_\mathrm{dw}$ and $\rho_\mathrm{v}$ denote the expected values of the critical defect densities at corresponding transitions.}
\end{figure}%
To extend the results to the full phase diagram (fig. \ref{fig2}), we show in fig. \ref{fig10} the dependence on $J_B/J$ of defect density values at the both transition temperatures. One can see that the domain wall density at $T=T_\mathrm{dw}$ has the universal value. The critical value of $\rho_\mathrm{v}(\bfs)$ at $T=T_\mathrm{v}$ tends to be universal while $T_\mathrm{dw}\leq T_\mathrm{v}$ (or $J_B\lesssim0.7J$), but when  $T_\mathrm{dw}> T_\mathrm{v}$ the density becomes larger its universal value, $\rho_\mathrm{v}=0.037(3)$ at $J_B=J$. $\rho_\mathrm{v}(\bfs)$ at $T=T_\mathrm{dw}$, $\rho_\mathrm{dw}$ at $T=T_\mathrm{v}$ as well as $\rho_\mathrm{v}(\sigma\bfs)$ at the both transition points is non-universal too.

It is interesting to note that the universality of the defect density at the multicritical (tetracritical) point $T_\mathrm{dw}=T_\mathrm{v}$ (see MP-point in fig. \ref{fig2}) implies that the multicritical behavior in the two-dimensional $\mathbb{Z}_2\oplus O(2)$ model is described by the decoupled Ising and $O(2)$ models.

\begin{figure}[t]
\center
\includegraphics[scale=0.45]{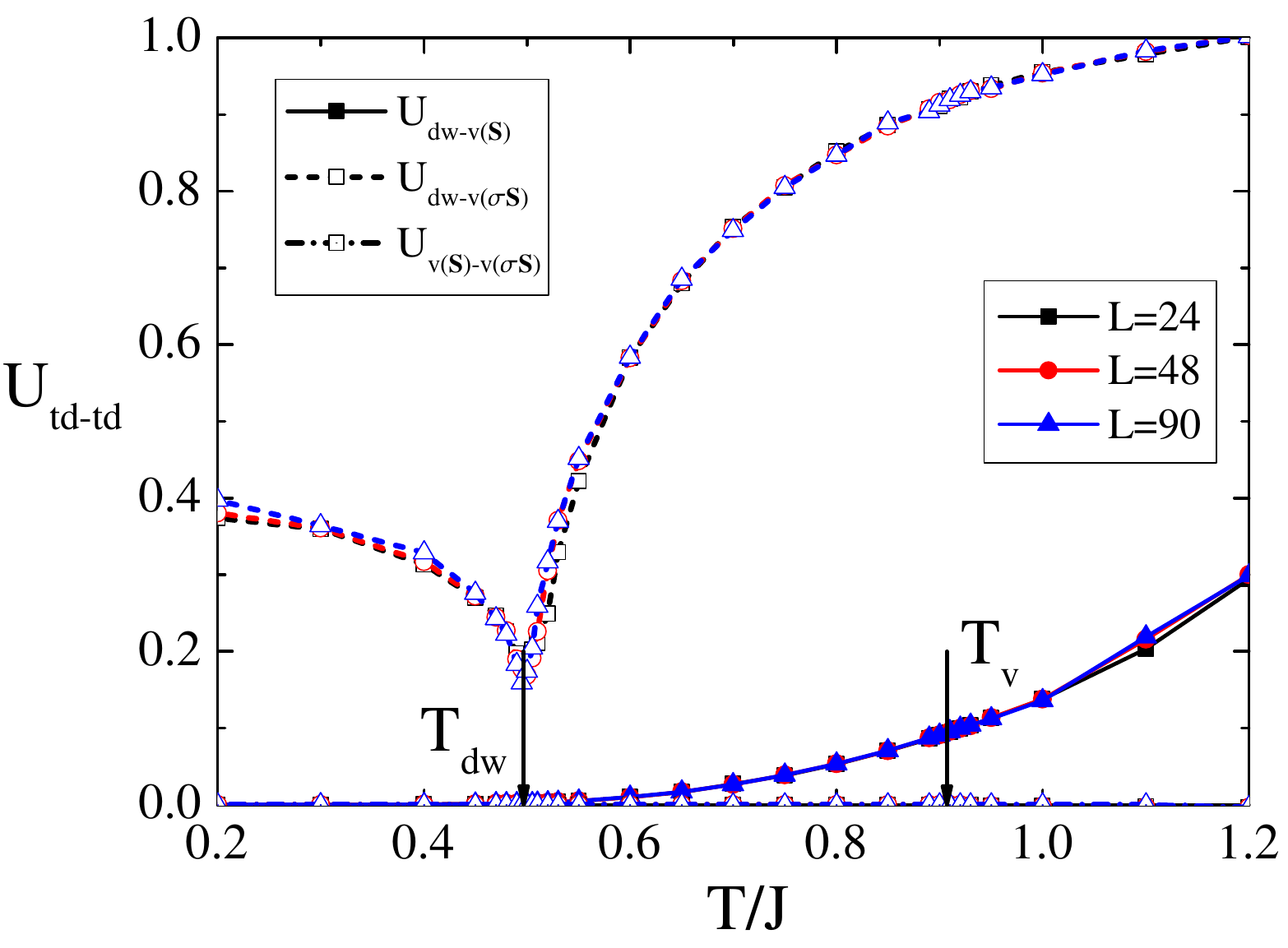}
\caption{\label{fig11} Defect-defect correlators in the Ising-XY model at $J_C=0$ and $J_B=0.25$.}
\end{figure}%
\begin{figure}[t]
\center
\includegraphics[scale=0.45]{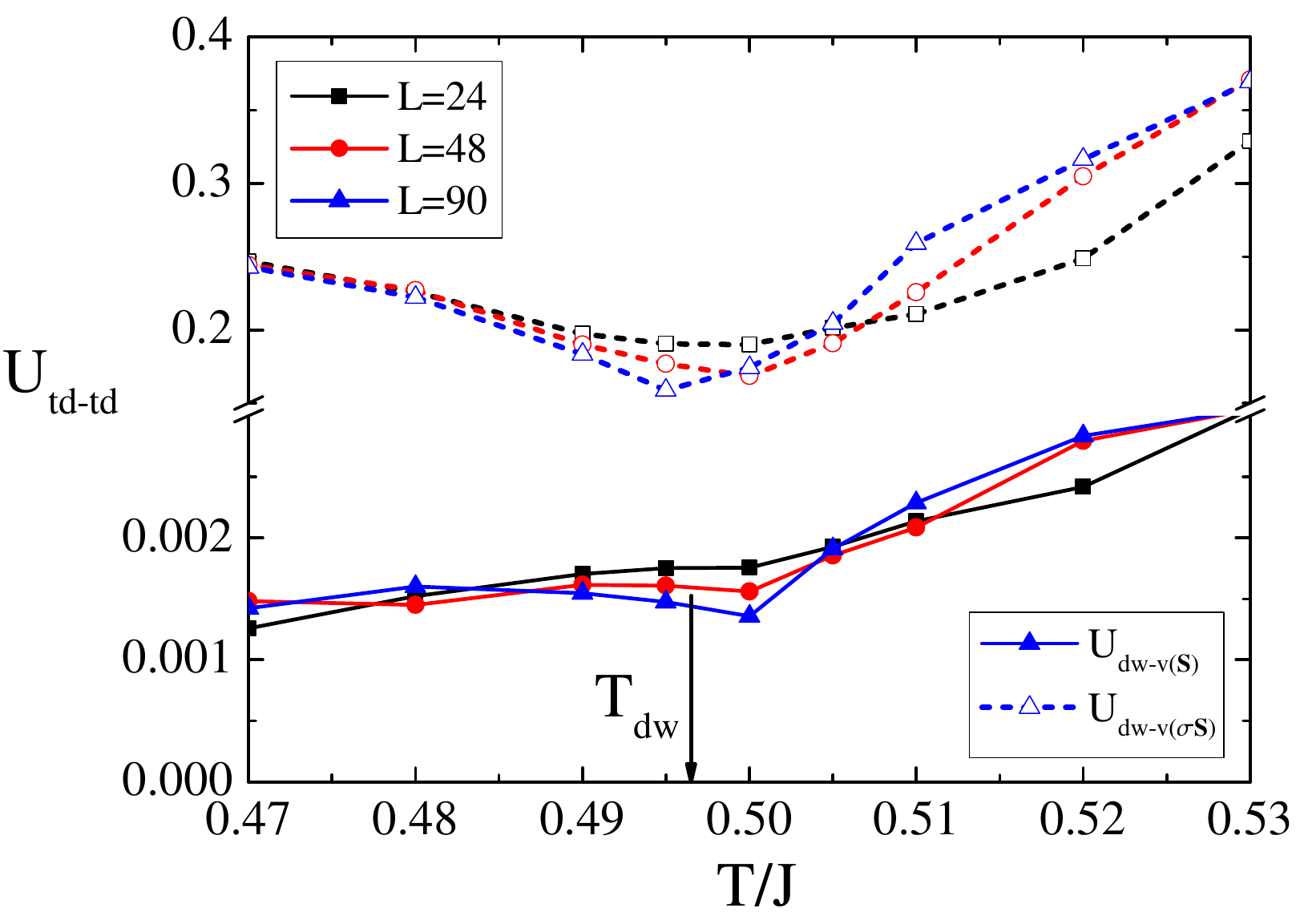}
\caption{\label{fig12} Wall-vortex correlator in the Ising-XY model at $J_C=0$ and $J_B=0.25$.}
\end{figure}%
\begin{figure}[t]
\center
\includegraphics[scale=0.45]{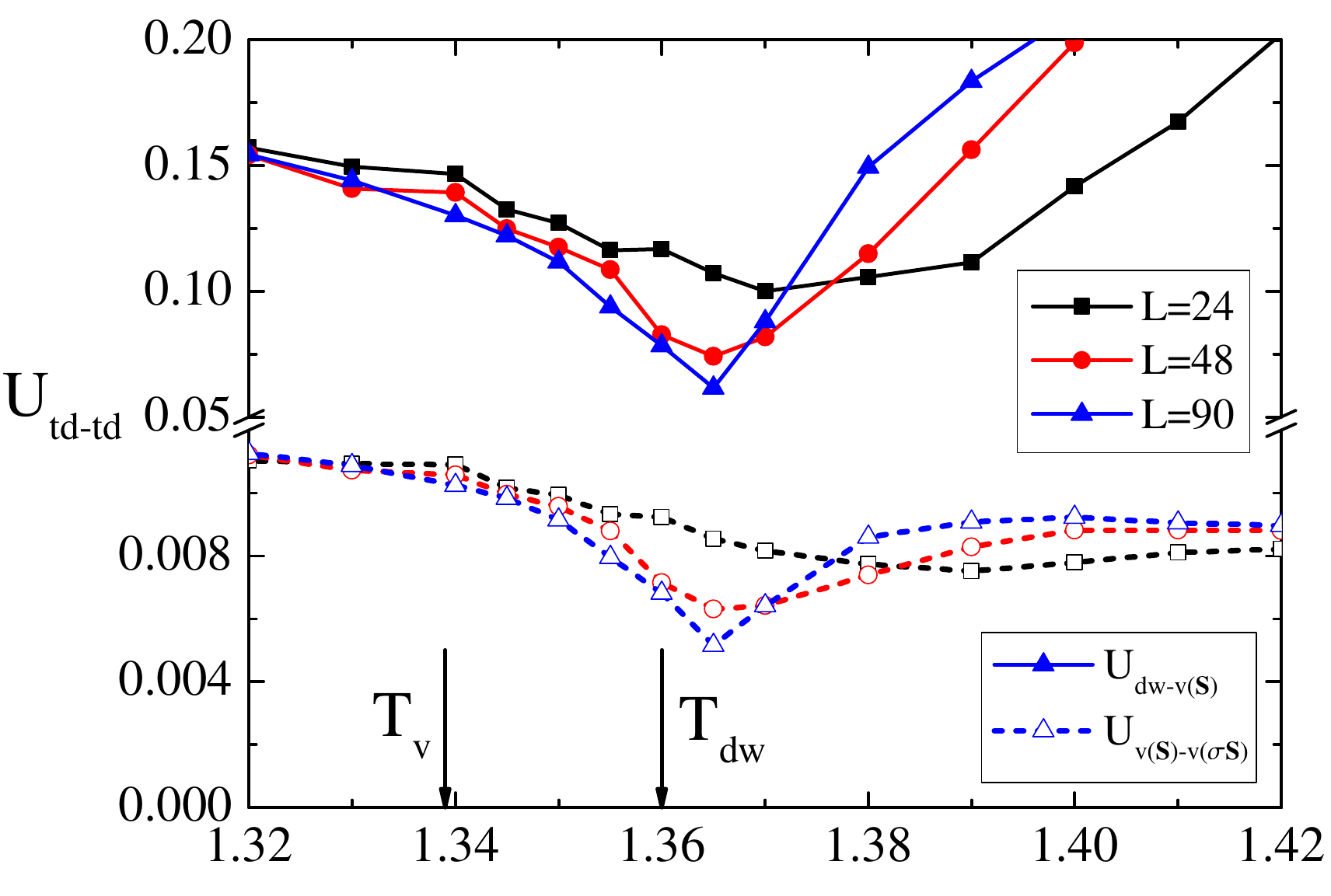}
\caption{\label{fig13} Defect-defect correlators in the Ising-XY model at $J_C=0$ and $J_B=1$.}
\end{figure}%
An interaction of domain walls with the $SO(2)$ sector of the Ising-XY model does not affect the critical behavior at the Ising transition point only if it vanishes at a large scale. It is confirmed by our simulations. Figs. \ref{fig11}-\ref{fig13} show that the wall-vortex correlator \eqref{td-td-corr} tends to zero with the lattice size increasing $L\to\infty$ at $T=T_\mathrm{dw}$. In more details, the wall-vortex correlators $U_{\mathrm{dw-v}(\bfs)}$ and $U_{\mathrm{dw-v}(\sigma\bfs)}$ remain non-zero with $L$ increasing at $T\neq T_\mathrm{dw}$. But at $T= T_\mathrm{dw}$ and $J_B\lesssim0.7J$ the correlators behave as
\begin{equation}
    U_{\mathrm{dw-v}(\bfs)}\sim (\ln L)^{-\frac12}, \quad
    T=T_\mathrm{dw},
\end{equation}
\begin{equation}
    U_{\mathrm{dw-v}(\sigma\bfs)}\sim (\ln L)^{-1}, \quad
    T=T_\mathrm{dw},
\end{equation}
and at $1\geq J_B\gtrsim0.7J$
\begin{equation}
    U_{\mathrm{dw-v}(\bfs)}\sim (\ln L)^{-1}, \quad
    T=T_\mathrm{dw}.
\end{equation}
The vortex-vortex correlator $U_{\mathrm{v}(\bfs)\mathrm{-v}(\sigma\bfs)}$ vanishes when $J_B\lesssim0.7J$ and at any temperature (fig. \ref{fig11}), and at $1\geq J_B\gtrsim0.7J$ it remains non-zero and behaves in the Ising transition point as (fig. \ref{fig13})
\begin{equation}
    U_{\mathrm{v}(\bfs)\mathrm{-v}(\sigma\bfs)}\sim (\ln L)^{-1}, \quad
    T=T_\mathrm{dw}.
\end{equation}

Of course, since the wall-vortex interaction \eqref{td-td-corr} is non-zero at the BKT transition temperature, the critical properties of vortices may differ from those in the pure $O(2)$ model in both cases $T_\mathrm{v}>T_\mathrm{dw}$ and $T_\mathrm{v}<T_\mathrm{dw}$. But at the multicritical point $T_\mathrm{v}=T_\mathrm{dw}$, the universality of vortex properties restores.

\subsection{Ising-$V_{3,2}$ model}

\begin{figure}[t]
\center
\includegraphics[scale=0.45]{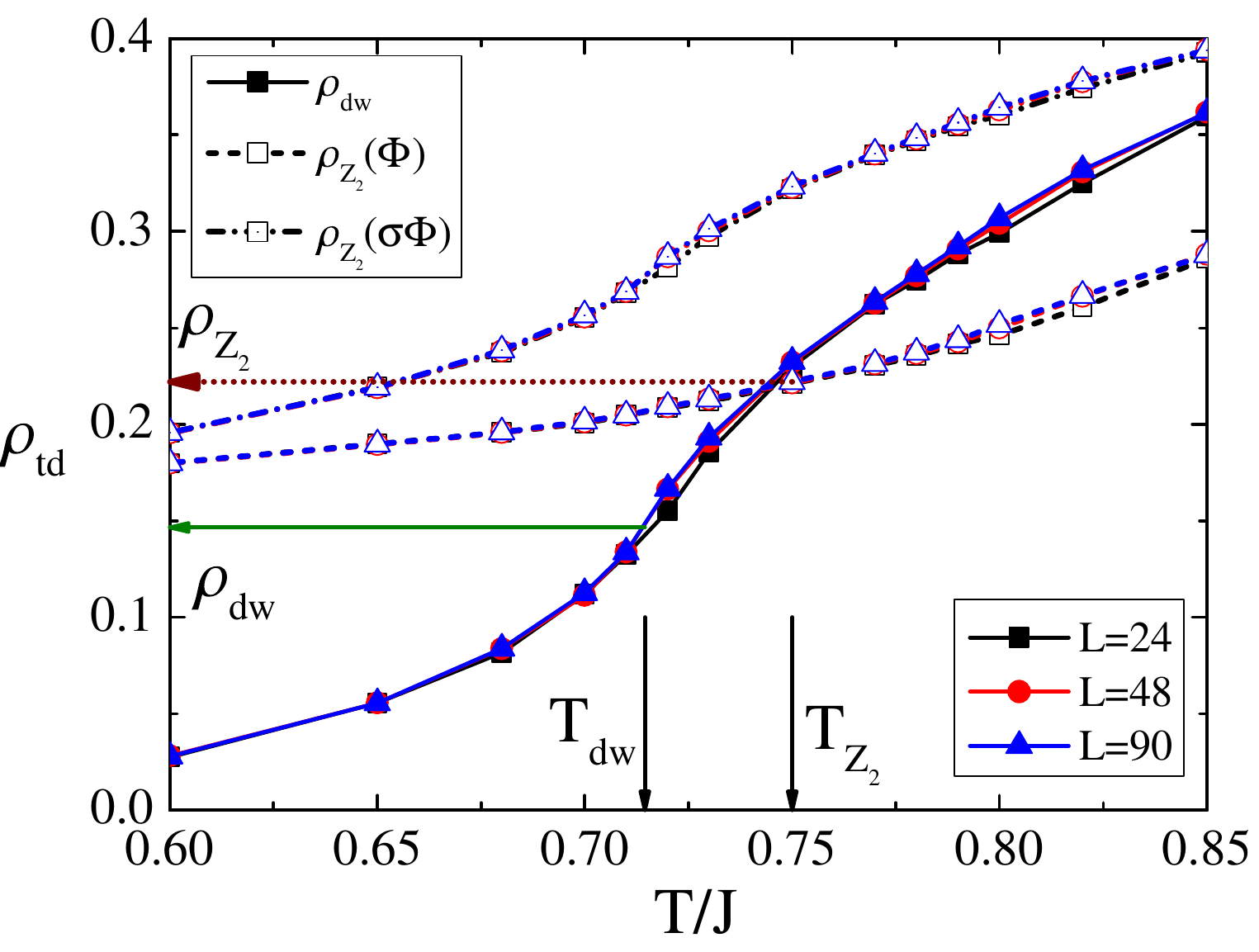}
\caption{\label{fig14} Densities of topological defects in the interpolating $V_{3,2}$-$V_{3,3}$ model at $J_B=0.5$. $\rho_\mathrm{dw}$ and $\rho_{\mathbb{Z}_2}$ denote the expected values of the critical defect densities at corresponding transitions.}
\end{figure}%
\begin{figure}[t]
\center
\includegraphics[scale=0.45]{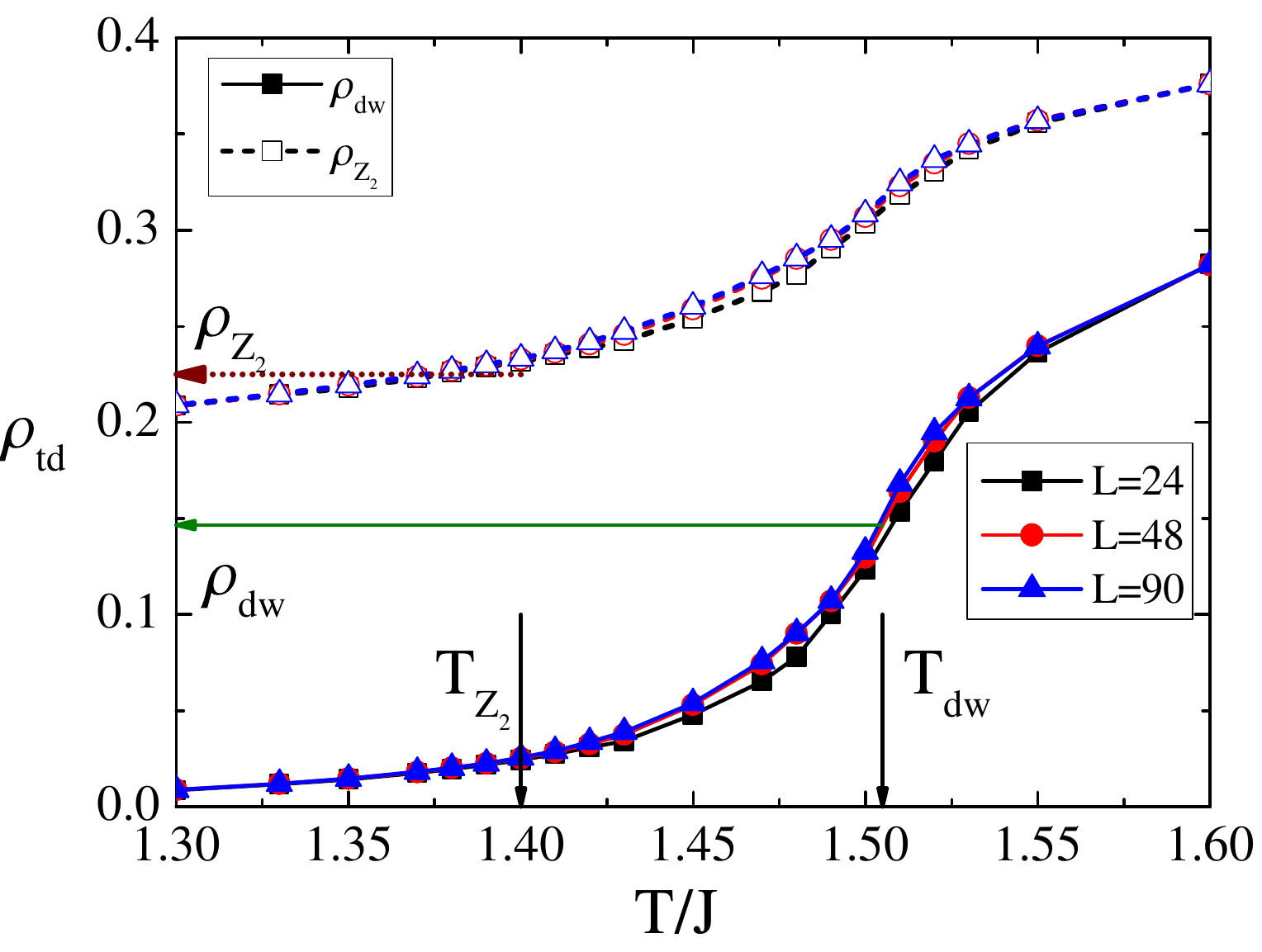}
\caption{\label{fig15} Densities of topological defects in the Ising-$V_{3,2}$ model at $J_B=1$. $\rho_\mathrm{dw}$ and $\rho_{\mathbb{Z}_2}$ denote the expected values of the critical defect densities at corresponding transitions.}
\end{figure}%
As we have discussed in ref.\cite{Sorokin18}, association of $\mathbb{Z}_2$ vortex-antivortex pairs increases ordering of a system, but it doesn't lead to appearance of long-range or quasi-long-range orders. Nevertheless, the increasing of temperature and concentration of $\mathbb{Z}_2$ vortices leads to a rather sharp change in the temperature behavior (crossover), which can be considered as an "almost critical"\ behavior. So we have assumed that the $\mathbb{Z}_2$ vortex density has an almost universal value at a crossover temperature, and have estimated the critical density as $\rho_{\mathbb{Z}_2}\approx0.225$.

\begin{figure}[t]
\center
\includegraphics[scale=0.45]{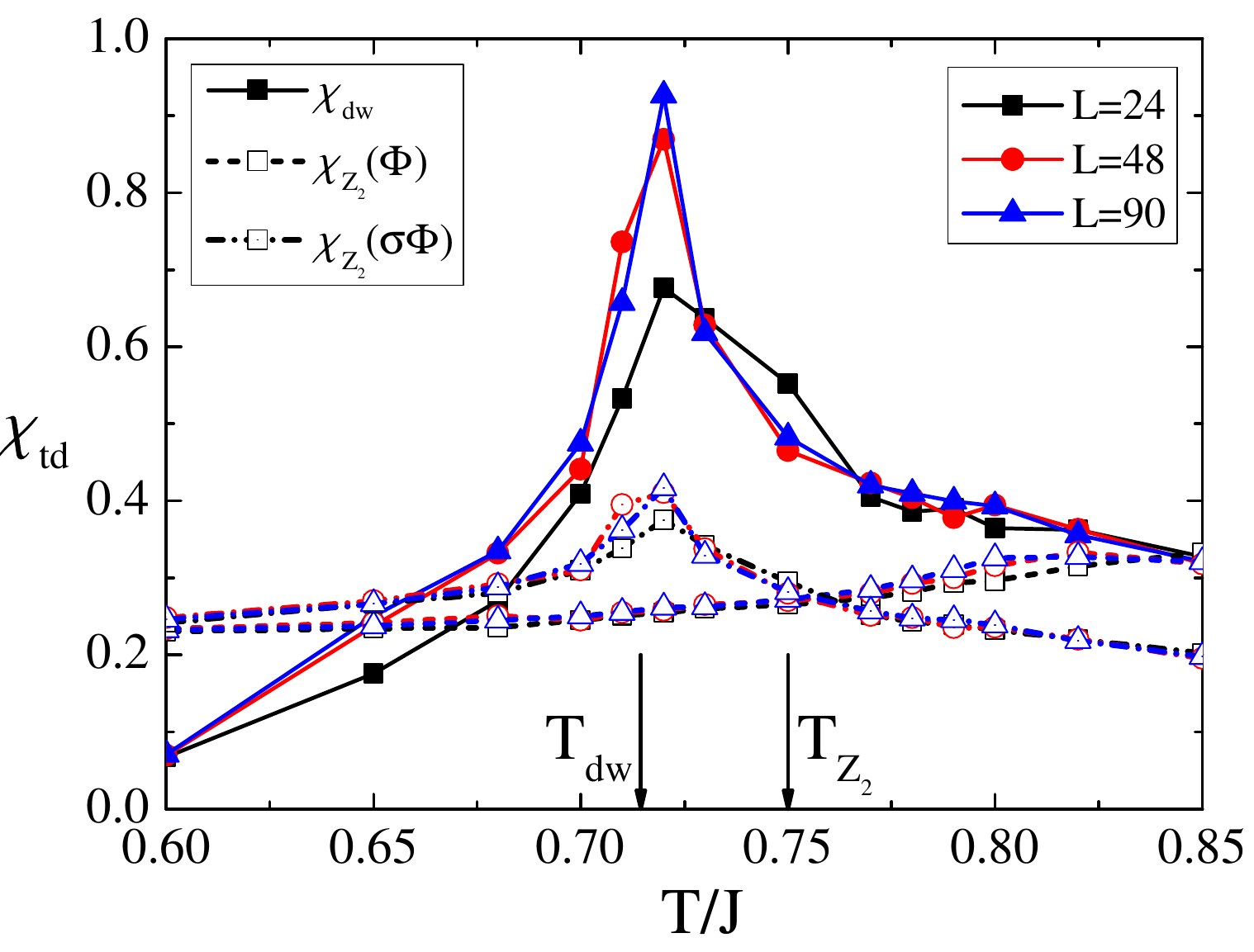}
\caption{\label{fig16} Topological susceptibilities in the interpolating $V_{3,2}$-$V_{3,3}$ model at $J_B=0.5$.}
\end{figure}

\begin{figure}[t]
\center
\includegraphics[scale=0.45]{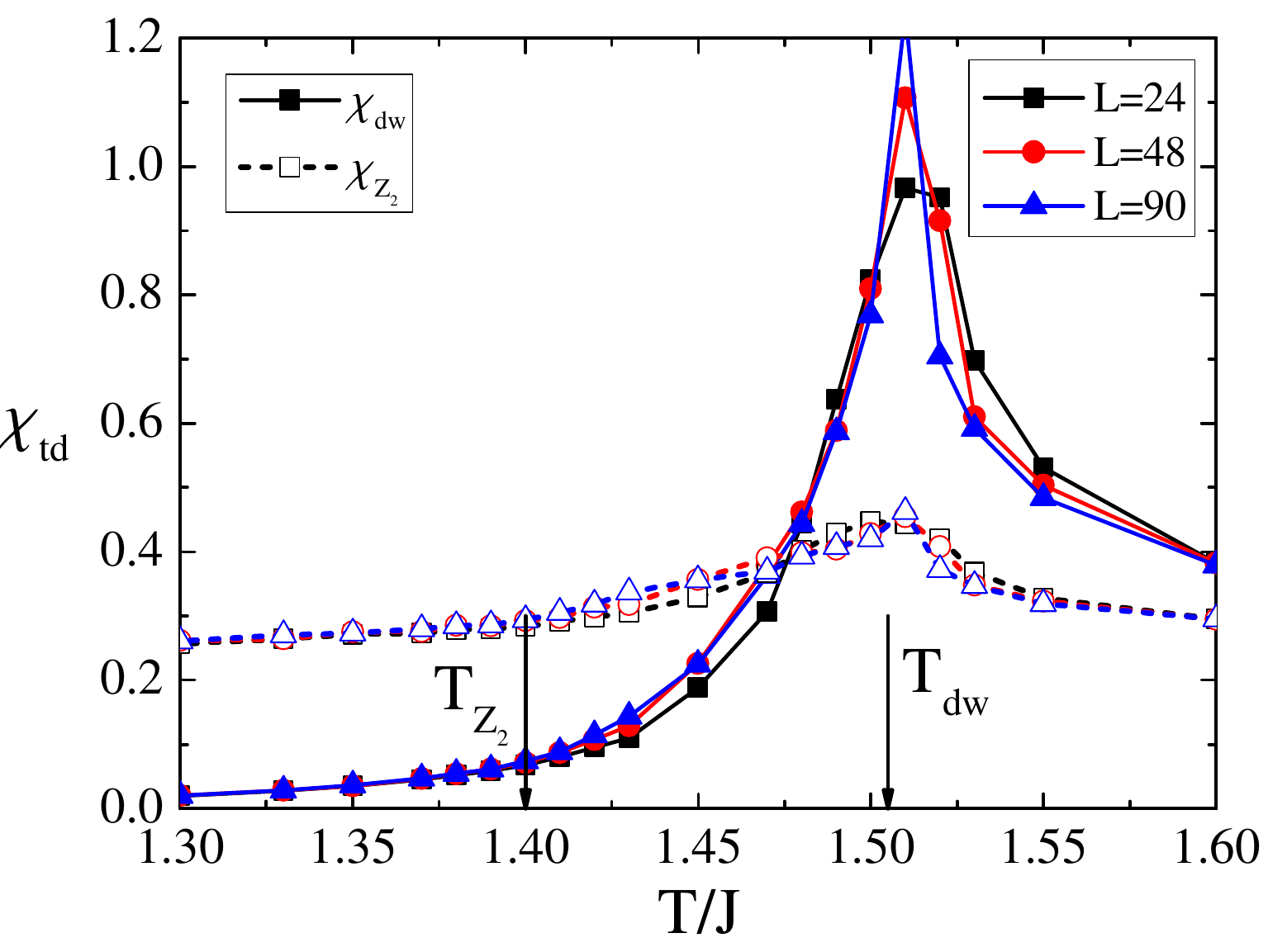}
\caption{\label{fig17} Topological susceptibilities in the Ising-$V_{3,2}$ model at $J_B=1$.}
\end{figure}

\begin{figure}[t]
\center
\includegraphics[scale=0.45]{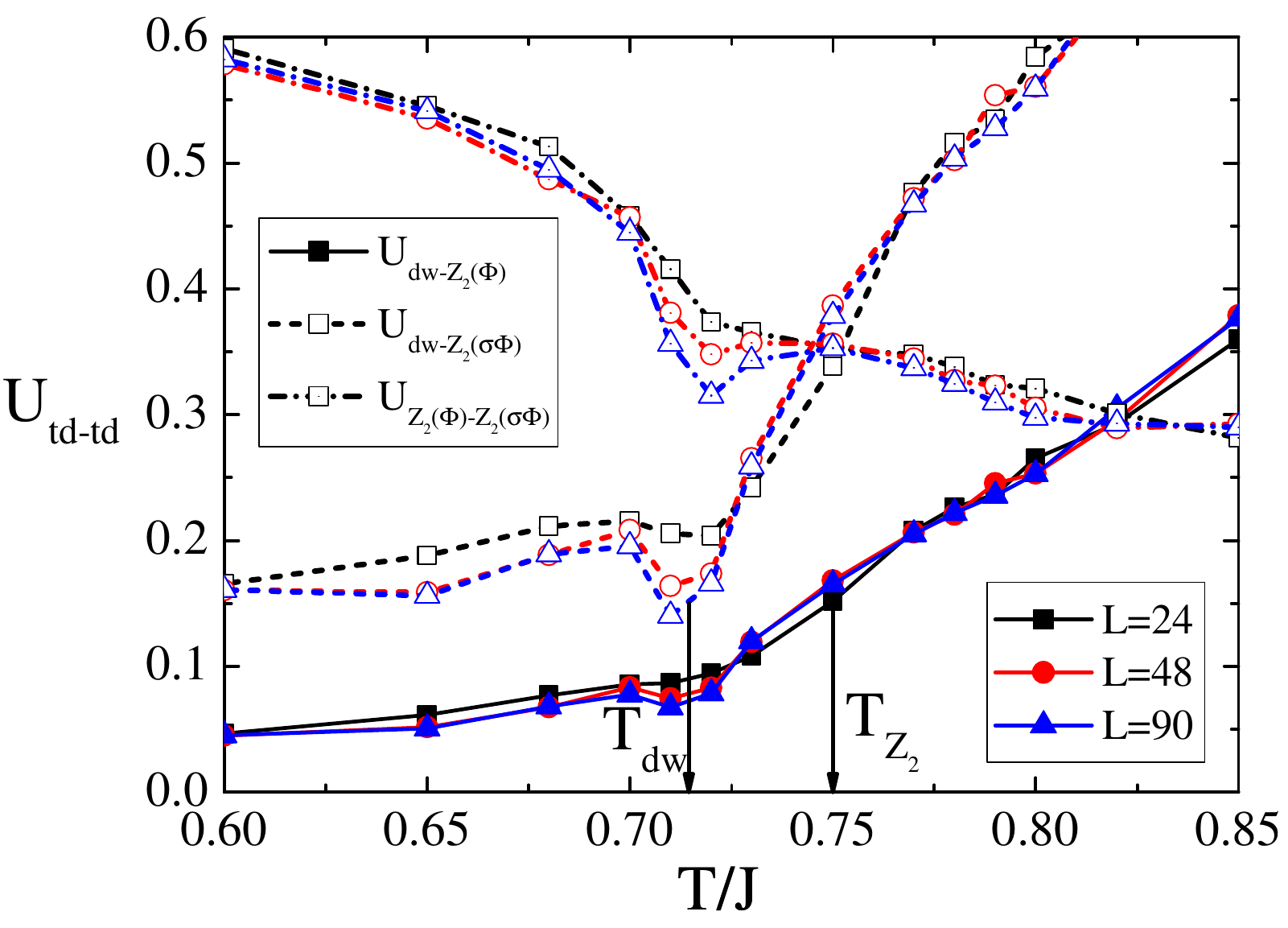}
\caption{\label{fig18} Defect-defect correlators in the interpolating $V_{3,2}$-$V_{3,3}$ model at $J_B=0.5$.}
\end{figure}

\begin{figure}[t]
\center
\includegraphics[scale=0.45]{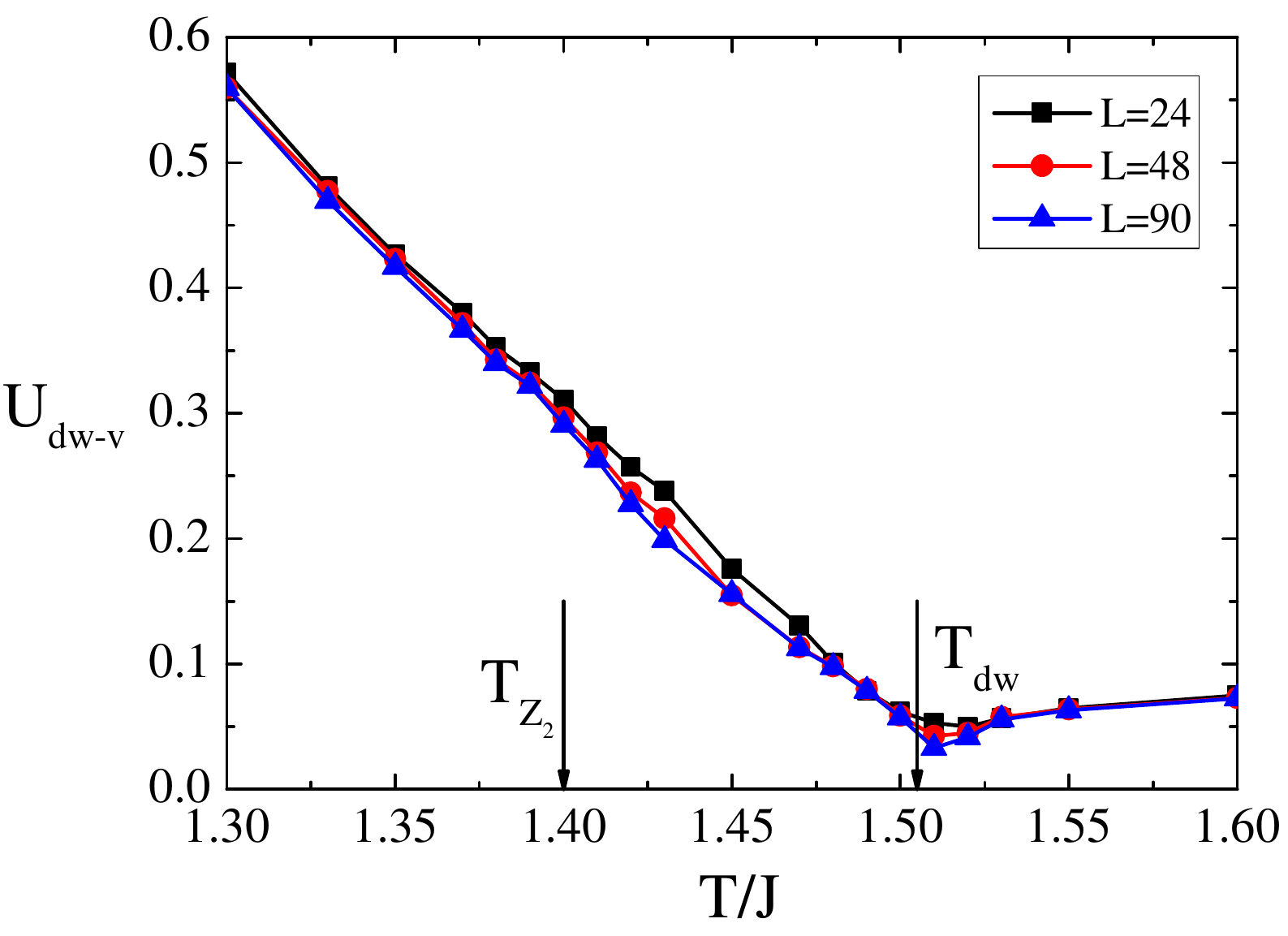}
\caption{\label{fig19} Defect-defect correlators in the Ising-$V_{3,2}$ model at $J_B=1$.}
\end{figure}

Although properties of $\mathbb{Z}_2$ vortices are significantly different from properties of ordinary vortices, the situation in the Ising-$V_{3,2}$ model is similar to one in the Ising-XY model (figs. \ref{fig14}-\ref{fig19}). In particular, when the Ising-like transition and crossover are well separated in temperature, the critical behavior is fully complies with the Ising and $V_{3,2}$ models, including the critical value of the topological defect densities (fig. \ref{fig14}). However, in the non-trivial case when domain walls induce the disorder in the continuous order parameter $J_B=J$ (fig. \ref{fig15}), the critical properties of domain walls and $\mathbb{Z}_2$ vortices are also close to the universal ones. So, as long as the critical behavior upon the transition in the discrete order parameter falls to the Ising universality class both above and below the multicritical point (see fig. \ref{fig3}), we assume that the multicritical behavior is described by the decoupled Ising and $V_{3,2}$ models.

Another similarity with the Ising-XY model is that the vortex density does not sense the Ising phase transition in the case of well-separated transition and crossover (figs. \ref{fig14} and \ref{fig16}). But in the case $J_B=J$, the $\mathbb{Z}_2$ vortex density has a singularity at $T=T_\mathrm{dw}$ (fig. \ref{fig17}).

Finally, we also observe that the wall-vortex interaction \eqref{td-td-corr} in the Ising-$V_{3,2}$ and interpolating $V_{3,2}$-$V_{3,3}$ models tends to zero value at the Ising transition point $T=T_\mathrm{dw}$ in the thermodynamical limit $L\to\infty$ (figs. \ref{fig18}, \ref{fig19}).

\section{Discussion}

In the paper \cite{Sorokin18} besides the hypothesis that the critical value of the topological defect density is universal upon a continuous phase transition, we have discussed several possibilities which may refute it. There is the case $J_B=J$ of the Ising-XY and Ising-$V_{3,2}$ models considered in this work among of them, when domain walls induce the disorder in the continuous order parameter and determine the sequence of phase transitions. In this work we show that this case confirms the hypothesis for a second order phase transition but not for a BKT transition.

\begin{figure}[t]
\includegraphics[scale=0.40]{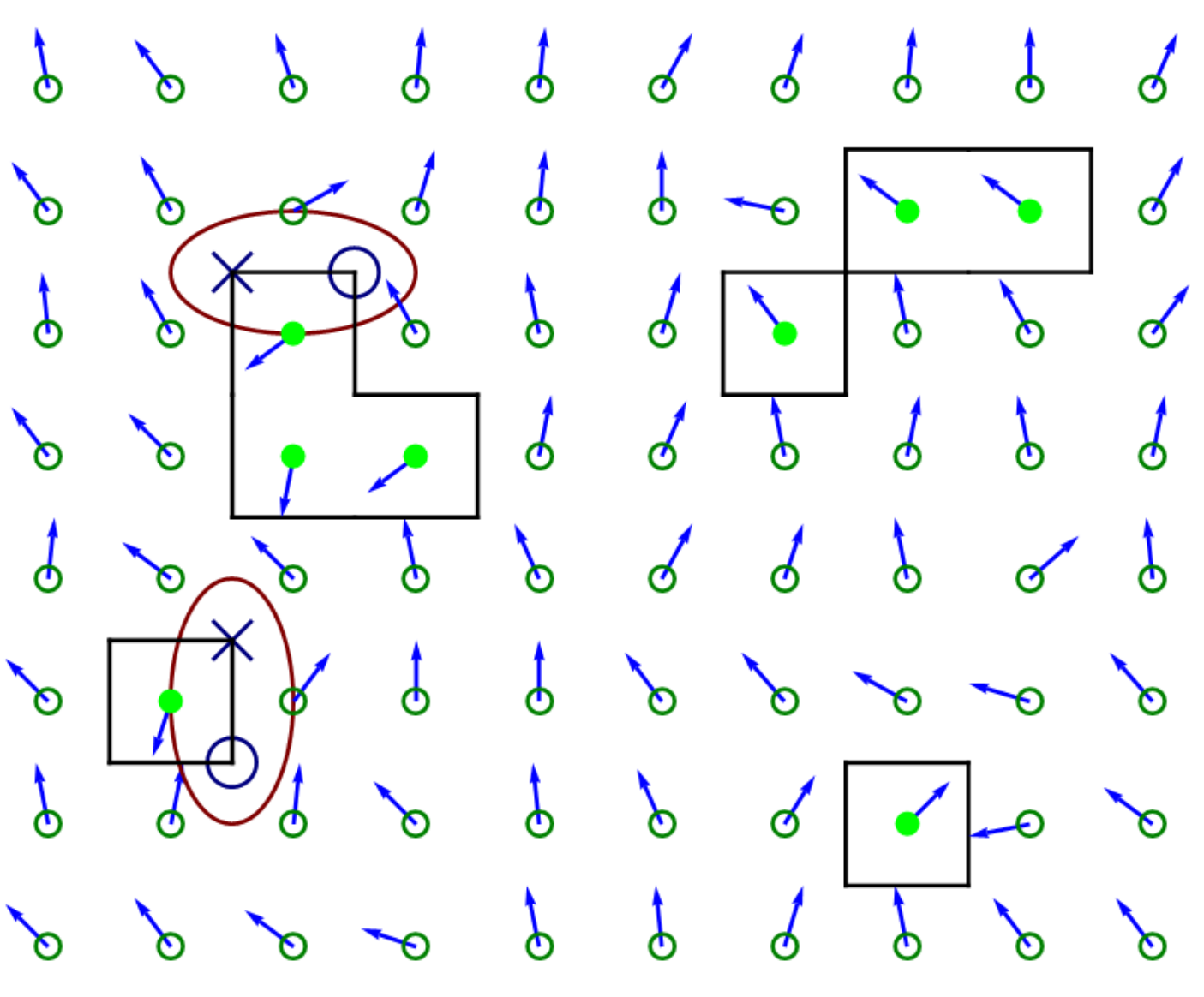}
\caption{\label{fig20}Shot of a simulation the Ising-XY model at $J_B=J$ and $T=T_\mathrm{v}$.}
\end{figure}%
Generally speaking, the defect-defect correlators \eqref{td-td-corr} denotes the possibility of a domain wall to lead to a dissociation of vortex-antivortex pair, or on the other hand, degree of conditionality of a domain wall configuration by a vortex configuration. So, in the case $U_{\mathrm{dw-v}}=1$ of the full correlation, any vortex (or antivortex) corresponds to a unit-length wall element and vice versa. The case $U_{\mathrm{dw-v}}=0$ means that vortex and wall configurations are independent $\langle\rho_\mathrm{dw}\rho_\mathrm{v}\rangle=\langle\rho_\mathrm{dw}\rangle\langle\rho_\mathrm{v}\rangle$. Any intermediate case $U_{\mathrm{dw-v}}>0$ means that the wall total length, its position on the lattice and non-equilibrium dynamics at least partly depend on a vortex configuration, the vortex and spin-wave dynamics. But we know that the critical properties of domain walls and their dynamics at the Ising critical point are determined by the conformal symmetry and the Schramm-Loewner evolution (see \cite{Cardy05} for a review). Therefore, the influence of the continuous sector of the Ising-XY (or Ising-$V_{3,2}$) model can either vanish at the Ising critical point, or violate the conformal invariance (a first-order transition), or change the universality class. In this work we find that the wall-vortex correlator vanishes at the Ising transition point in accordance with the first scenario. The second scenario may realized too in the Ising-XY model \cite{Granato87,Granato91} as well as in the Ising-$V_{3,2}$ model \cite{Sorokin18}. We do not know arguments in favor of the possibility for the third scenario realization.

The Kosterlitz-Thouless renormalization group approach \cite{Kosterlitz73,Kosterlitz74} does not impose restrictions on the value of the total vortex density calculated in this work. The theory predicts that the density of {\it free} vortices (not associated in pairs) behaves above a BKT transition as
\begin{equation}
 \rho_\mathrm{free}\sim \xi^{-2}\sim\exp\left(2b\sqrt{\frac{T-T_\mathrm{v}}{T_\mathrm{v}}}\right),
\end{equation}
where $\xi$ is the correlation length, $b$ is some constant (see, e.g. ref. \cite{Korshunov06}). So, the non-vanishing wall-vortex correlator at a BKT transition does not lead to a contradiction with the predictions of the theory, but explains the larger value of the vortex density in the $J_B=J$ case. To illustrate said above, we show a random shot of a simulation in fig. \ref{fig20}. One can see that vortices induced by domain walls are associated in pairs. Such vortices contribute to the total vortex density, but are excluded from the calculation of the free vortex density.

\begin{acknowledgments}
This work is supported by the RFBR grant No 16-32-60143.
\end{acknowledgments}

\end{document}